\documentclass[9pt, sigconf]{acmart}

\usepackage{balance}

\def\BibTeX{{\rm B\kern-.05em{\sc i\kern-.025em b}\kern-.08emT\kern-.1667em\lower.7ex\hbox{E}\kern-.125emX}}
    
\copyrightyear{2020}
\acmYear{2020}
\setcopyright{acmcopyright}
\acmConference[SIGMOD '20] {2020 ACM SIGMOD International Conference on Management of Data}{June 14--19, 2020}{Portland, OR, USA}
\acmBooktitle{2020 ACM SIGMOD International Conference on Management of Data (SIGMOD'20), June 14--19, 2020, Portland, OR, USA}
\acmPrice{15.00}
\acmDOI{10.1145/XXXXXX.XXXXXX}
\acmISBN{978-1-4503-6735-6/20/06}

\newcommand{\eat}[1]{}

\usepackage{array}
\usepackage[figuresright]{rotating}
\usepackage{multirow}
\usepackage{footnote}
\usepackage{listings}
\usepackage{color}
\usepackage{url}
\usepackage{comment}
\usepackage{algorithm, algpseudocode}
\usepackage{subcaption}
\usepackage{amsmath, amssymb}
\usepackage{placeins, float}
\usepackage{tabularx}
\usepackage{pgfplots, pgfplotstable}
\usepackage{flushend}
\usepackage{pifont}

\pgfplotsset{compat=1.14, width=4.3cm}
\usetikzlibrary{patterns}
\usepgfplotslibrary{statistics}

\DeclareMathOperator*{\argmax}{arg\,max}
\DeclareMathOperator*{\argmin}{arg\,min}
\definecolor{dkgreen}{rgb}{0,0.6,0}
\definecolor{gray}{rgb}{0.5,0.5,0.5}
\definecolor{mauve}{rgb}{0.58,0,0.82}

\newcommand{\mkcomment}[1]{}

\newcommand{\relm}{{\sf RelM}}

\newcommand{\poolheap}{{\sf Heap}}
\newcommand{\poolcache}{{\sf Cache Storage}}
\newcommand{\poolshuffle}{{\sf Task Shuffle}}
\newcommand{\poolunmgd}{{\sf Task Unmanaged}}
\newcommand{\poolcode}{{\sf Code Overhead}}
\newcommand{\poolold}{{\sf Old}}
\newcommand{\pooleden}{{\sf Eden}}

\newcommand{\sort}{{\sf SortByKey}}
\newcommand{\wc}{{\sf WordCount}}
\newcommand{\kmeans}{{\sf K-means}}
\newcommand{\pagerank}{{\sf PageRank}}
\newcommand{\svm}{{\sf SVM}}

\newcommand{\rmv}{{\sf U}}
\newcommand{\maximizeresource}{{\em MaxResourceAllocation}}
\newcommand{\exhaustive}{{\em Exhaustive Search}}
\newcommand{\gaussian}{{\sf BO}}
\newcommand{\randomforest}{{\sf RF}}
\newcommand{\optgaussian}{{\sf GBO}}
\newcommand{\optrandomforest}{{\sf RF$^{fast}$}}
\newcommand{\guided}{{\sf GBO}}
\newcommand{\bayesian}{{\sf BO}}
\newcommand{\reinforce}{{\sf DDPG}}

\newcommand{\mc}[1]{{\mathcal #1}}
\newcommand{\mb}[1]{{\mathbf #1}}

\newcommand{\squishlist}{
 \begin{list}{$\bullet$}
  { \setlength{\itemsep}{0pt}
     \setlength{\parsep}{1pt}
     \setlength{\topsep}{1pt}
     \setlength{\partopsep}{0pt}
     \setlength{\leftmargin}{1.5em}
     \setlength{\labelwidth}{1.5em}
     \setlength{\labelsep}{0.5em} } }

\newcommand{\squishlisttwo}{
 \begin{list}{$\number$}
  { \setlength{\itemsep}{0pt}
    \setlength{\parsep}{0pt}
    \setlength{\topsep}{0pt}
    \setlength{\partopsep}{0pt}
    \setlength{\leftmargin}{1em}
    \setlength{\labelwidth}{1em}
    \setlength{\labelsep}{0.5em} } }

\newcommand{\squishend}{
  \end{list}  }

\begin{document}

\date{}

\title{Black or White? How to Develop an AutoTuner for Memory-based Analytics [Extended Version]}\titlenote{This research is supported by NSF grant IIS-1423124.}

\author{Mayuresh Kunjir}
\affiliation{%
	 \institution{Duke University}
}
\email{mayuresh@cs.duke.edu}

\author{Shivnath Babu}
\affiliation{%
  \institution{Unravel Data Systems}
}
\email{shivnath@unraveldata.com}

\renewcommand{\shortauthors}{Kunjir and Babu}

\begin{abstract}

There is a lot of interest today in building autonomous (or, self-driving) data processing systems. An emerging school of thought is to leverage AI-driven ``black box" algorithms for this purpose. In this paper, we present a contrarian view. We study the problem of autotuning the memory allocation for applications running on modern distributed data processing systems. 
For this problem, we show that an empirically-driven ``white-box" algorithm, called \relm, that we have developed provides a {\em close-to-optimal} tuning at a fraction of the overheads compared to state-of-the-art AI-driven ``black box" algorithms, namely, Bayesian Optimization (BO) and Deep Distributed Policy Gradient (DDPG).
The main reason for \relm's superior performance is that the memory management in modern memory-based data analytics systems is an interplay of algorithms at multiple levels: (i) at the resource-management level across various containers allocated by resource managers like Kubernetes and YARN, (ii) at the container level among the OS, pods, and processes such as the Java Virtual Machine (JVM), (iii) at the application level for caching, aggregation, data shuffles, and application data structures, and (iv) at the JVM level across various pools such as the Young and Old Generation. \relm\ understands these interactions and uses them in building an analytical solution to autotune the memory management knobs. In another contribution, called \guided, we use the \relm's analytical models to speed up Bayesian Optimization.
Through an evaluation based on  Apache Spark, we showcase that \relm's  recommendations are significantly better than what commonly-used Spark deployments provide, and 
are close to the ones obtained by  brute-force exploration; while {\guided} provides optimality guarantees for a higher, but still significantly lower compared to the state-of-the-art AI-driven policies, cost overhead.

\end{abstract}



\maketitle

\vspace{-4mm} 
\section{Introduction}
\label{sec:intro}

Modern data analytics systems, e.g. Spark, Tez, and Flink, are increasingly using memory both for data storage and fast computations. However, memory is a limited resource that must be managed carefully 
by three players: 

\squishlist
\item {\em Application Developer}:
 Judging by the magnitude of StackOverflow posts and user surveys~\cite{pepperdata, 123errors}, `out-of-memory' errors is a major cause of unreliable application performance. To safeguard against such errors, developers need an understanding of how much memory their application really needs and how to set the appropriate memory configurations. The  prevalent rule-of-thumb to ``throw more memory at your applications'' is not the best approach while considering costs or the interests of other users. 
\eat{
A few heuristics-based solutions can be used~\cite{ryza, thoth-action}, but they require an expert knowledge of the underlying application platform. 
}

\item {\em Resource Manager}:
A resource manager in a multi-tenant setting, e.g.,  
YARN, needs to carefully allocate resources to meet the application performance goals of multiple tenants. Over-allocation leads to wasted resources and a lower throughput, while under-allocation could mean higher latency for tenants. Both problems are commonly observed in production clusters~\cite{tempo, rayon, robus}.  

\item {\em Application Platform}:
The onus of ensuring a {\em safe} usage of memory is predominantly on the application platforms. Memory is used for various operations such as joins/aggregation, caching inputs/intermediate results, data shuffling/repartitioning, and sending intermediate/output data over network. Arbitrating memory across these operations is critical in ensuring a reliable and fast execution. Improving memory management is a major focus in modern analytics platforms~\cite{tungsten, flink}.

\squishend 

\noindent{\bf Challenges and Contributions:}

A major challenge faced by the data processing platforms arises from the fact that 
the memory management decisions in the data processing platforms are made at multiple levels (viz. the resource-management level, at the container level, at the application level, and inside the Java Virtual Machine) with complex interplays involved amongst the decisions and the performance metrics.
Data analytics applications vary widely in terms of both the computational model (e.g., SQL, shuffling, iterative processing) and the physical design of input data (e.g., partition sizes) translating to huge variations in their resource consumption patterns.
Consequently, they exhibit complex response surfaces to configuration options relating to resource usage~\cite{ituned, bestconfig}. 
Section~\ref{sec:interactions} presents a detailed empirical analysis showing the impact and interactions of memory management options to further press this point. It is found that the default settings provided by the commonly-used system deployments leave a lot of room for improvement in terms of the reliability and the running time of the applications. Users running the applications on such deployments desire an automated tuning solution that recommends better memory configurations for their workload in a short span of time. Building such solutions is the focus of this paper.

The workload we consider is a data analytics application workflow along with its input data. Given the wide variety in the possible computational patterns and the physical design of data, building analytical cost-based performance models is non-trivial. 
Much of the previous work has focussed on training performance models {\em offline}, using a small-scale benchmark test bed, historical performance data, or from application performance under low workload~\cite{ernest, ottertune, cart, cdbtune}. Offline training poses two difficulties in applying the models in real-world settings: (i) Experiments on small-scale test beds may not represent intricacies of real applications accurately; and (ii) Applying models in a changed environment or workload may involve an expensive online learning cycle. 

One option for tuning is {\em online} model-free exploration of the configuration space, typically involving a combination of random sampling and local search
~\cite{smarthill, mronline, elastisizer, bdaf, bestconfig}. However, this black-box approach can be very expensive given the complex non-linear response surfaces and the high costs associated with running each experiment.

Speeding up exploration calls for an improvement-based policy which follows a Sequential Model-based Optimization (SMBO) approach~\cite{smac}. SMBO iterates between fitting a surrogate model and using it to recommend the next probe of configuration space.
Bayesian optimization (\bayesian)~\cite{bayesian-book} is a powerful state-of-the-art SMBO technique that 
provides a theoretically-justified exploration of the configuration space with improvement guarantees. 
Another exciting possibility is to use a deep reinforcement learning approach that uses a reward-feedback approach to tuning. Deep Distributed Policy Gradient (\reinforce)~\cite{ddpg} is a powerful technique providing a model-free, actor-critic algorithm which can operate on continuous action (configuration) spaces.

We approach the tuning problem by developing a deep understanding of the internal memory management options.
Rather than directly modeling the high level tuning objectives, such as latency, we model the impact of the memory configurations on the efficiency of the system resource utilization and the reliability of execution. This understanding is used to develop an algorithm, called \relm, that quickly tunes the memory management options using a very small number (one or two) of profiled application runs. At the core of \relm\ is a set of simple analytical models that estimates the requirements of the various competing memory pools within an application. Using the models, \relm\ guarantees a {\em safe}, that is, free of out-of-memory errors and, simultaneously, highly resource-efficient configuration.

In another contribution, we use \relm's analytical models to speed up the black-box tuning of \bayesian. This modification, called {\em Guided Bayesian Optimization} (\guided), plugs in metrics derived from an application profile relating to reliability, efficiency, and performance overheads to the \bayesian\ model. 


The two solutions we have designed for tuning memory management decisions in data analytics both improve the state-of-the-art significantly and also present an interesting trade-off to the end user: While \relm\ offers a {\em good} (performing within top 5 percentile of the exhaustively searched configurations) tuning recommendation with a minimal training overhead, \guided\ guarantees optimality given an allowance for a slightly higher overhead. The reinforcement learning approach (\reinforce) is shown to possess a great ability to adapt to high dimensional spaces as well as to changes in the test environment thereby making a strong case for use in other related auto-tuning problems.

\begin{figure}
\centering
\vspace*{-2mm}
 \includegraphics[width=0.6\columnwidth]{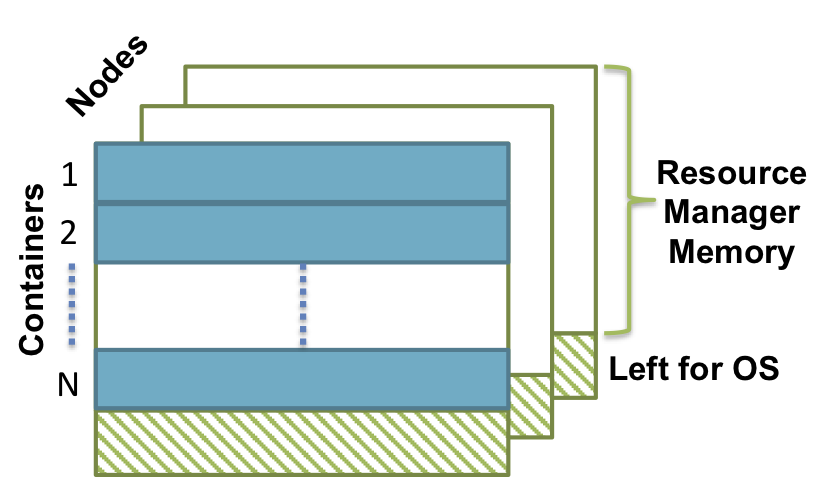}
\vspace*{-4mm}
 \caption{Memory managed by Resource Manager}
 \label{fig:rmlevel}
\quad
 \includegraphics[width=0.9\columnwidth]{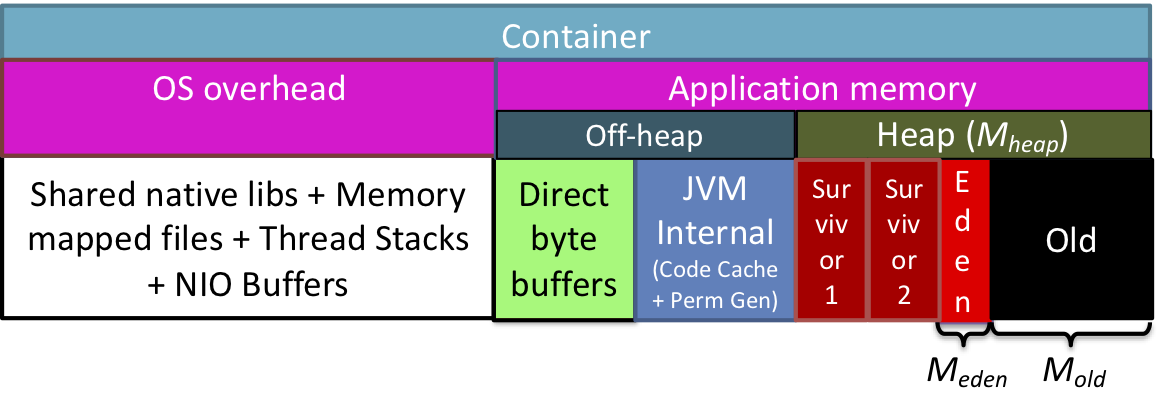}
\vspace*{-6mm}
 \caption{Container memory managed by JVM}
 \label{fig:jvmlevel}
\quad
 \includegraphics[width=0.8\columnwidth]{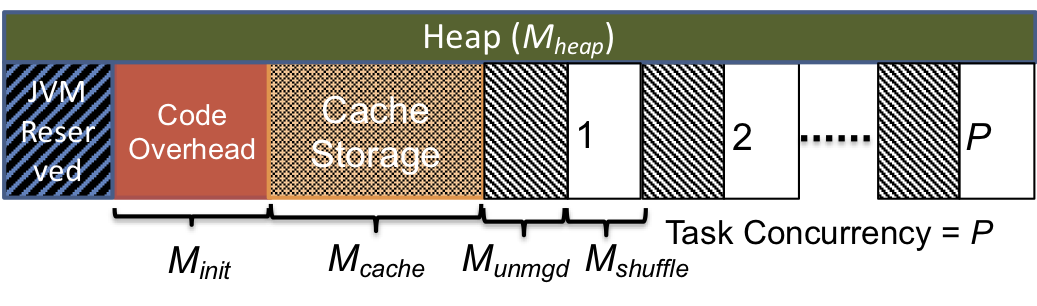}
\vspace*{-6mm}
 \caption{Heap managed by application framework}
\vspace*{-8mm}
 \label{fig:applevel}
\end{figure}

\vspace*{-3mm}
\section{Problem Overview}
\label{sec:problem}
\vspace*{-1mm}
\subsection{Memory-based Analytics}
\label{sec:mem-overview}
\vspace*{-1mm}

\noindent Data analytics clusters employ a resource manager, such as Yarn~\cite{yarn}, to allocate cluster resources to applications. Each application is provided with a set of {\em containers} by the resource manager. A container is simply a slice of physical resources carved out of a node allocated exclusively to the application. Figure~\ref{fig:rmlevel} shows how cluster memory is allocated to multiple containers. Many popular data analytics systems (e.g., Spark, Flink, and Tez) use a JVM-based architecture for memory management. For applications running on these systems, a JVM process is executed inside every container allocated by the resource manager. As shown in Figure~\ref{fig:jvmlevel}, the container memory is divided into two parts: (a) Memory available to the JVM process, and (b) An overhead space used by the operating system for process management. The JVM further divides its allocation into a heap space and an off-heap space. All objects, except native byte buffers, created by the application code are allocated on the heap and are managed by the JVM's generational heap management as described next.

One of the most salient features of the JVM is the {\em safety} in memory management. 
Unlike in the native programming languages, 
applications written in JVM languages do not explicitly allocate and free memory. Instead, the JVM periodically runs a process of garbage collection (GC in short) that frees up unreferenced objects from \poolheap. \poolheap\ is internally organized into multiple pools, each holding objects of certain {\em age}. The number of pools and their sizes are determined by a GC policy configured at the time of process launch. We focus on the default policy, called ParallelGC. ParallelGC uses two pools: Young generation and Old generation. As the names suggest, the Young pool stores newly created objects and the Old pool stores long-living objects. 

ParallelGC splits up the young generation into one {\em Eden} space and two {\em Survivor} spaces only one of which is occupied at any given time. Each of the pools is a contiguous block in memory. Newly created objects go to \pooleden\ first. When \pooleden\ is filled up, a collection called {\em Young GC} is triggered to collect unreferenced objects from \pooleden\ and the occupied Survivor. Objects that have {\em aged enough} (determined by GC parameters `InitialTenuringThreshold' and `MaxTenuringThreshold'~\cite{java-docs}) are moved to the \poolold\ pool while the other objects go to the other empty survivor. When a Young GC process finds an almost full old generation, it triggers a {\em Full GC} process which collects all unreferenced objects from \poolold, moves surviving objects from Young to \poolold, and compacts the \poolold\ pool. 

Certain phases of any GC process includes {\em stop-the-world} pauses where the application threads need to be suspended. 
Minimizing these pauses require parameter tuning. Key tuning options controlling ParallelGC are related to sizing the young and old pools. Parameter {\em NewRatio} sets the ratio of the capacity of \poolold\ to the capacity of Young. The capacity of \pooleden\ within Young is decided by parameter {\em SurvivorRatio} which gives the ratio of the capacity of \pooleden\ to the capacity of a Survivor space.

Figure~\ref{fig:applevel} shows how \poolheap\ is organized into different pools from the application's perspective. Except the space reserved for the JVM's internal objects and a survivor space, the entire \poolheap\ is used by application inputs and code objects. This space can be broadly categorized into three pools: 
\squishlisttwo
\item[1] {\em Code Overhead}: Memory required for application code objects ($M_{i}$). Treated as a constant overhead.
\item[2] {\em Cache Storage}: Memory used to store the data cached by application ($M_{c}$). In particular, storing intermediate results in memory is beneficial during iterative computations.
\item[3] {\em Task Memory}: The rest of the memory is used by application tasks. The number of tasks running concurrently is set as a configuration parameter, Task Concurrency, which determines the share of memory each task gets to use. Each task needs memory for: (a) Shuffle processing tasks such as sort and aggregation ($M_{s}$), (b) Input data objects and serialization/deserialization buffers ($M_{u}$).
\squishend 

\begin{table}
\centering
\caption{Parameters controlling memory pools across multiple levels: Container, Application Framework, and JVM displayed in order from top to bottom.}
\vspace*{-4mm}
\resizebox{\columnwidth}{!}{%
 \begin{tabular}{|c c c|}
 \hline
 Parameter & Description & Pool(s) controlled\\
 \hline
 {\bf Heap Size} & \begin{tabular}{@{}c@{}}Heap size in a container\end{tabular} & \begin{tabular}{@{}c@{}}Heap ($M_{h}$)\end{tabular} \\
 \hline
 {\bf Cache Capacity} & \begin{tabular}{@{}c@{}}Cache storage as a\\ fraction of Heap\end{tabular} & \begin{tabular}{@{}c@{}}Cache Storage ($M_{c}$)\end{tabular} \\
 {\bf Shuffle Capacity} & \begin{tabular}{@{}c@{}}Shuffle memory as\\ a fraction of Heap\end{tabular} & \begin{tabular}{@{}c@{}}Task Shuffle ($M_{s}$)\end{tabular}\\
 \begin{tabular}{@{}c@{}}{\bf Task} {\bf Concurrency}\end{tabular} & \begin{tabular}{@{}c@{}}Number of tasks running\\ concurrently\end{tabular} & \begin{tabular}{@{}c@{}}Task Unmanaged ($M_{u}$)\end{tabular}\\
 \hline
 {\bf NewRatio} & \begin{tabular}{@{}c@{}}Ratio of Old capacity\\ to Young capacity\end{tabular} & \begin{tabular}{@{}c@{}}Old ($M_{o}$)\end{tabular} \\
 {\bf SurvivorRatio} & \begin{tabular}{@{}c@{}}Ratio of Eden capacity\\ to Survivor space\end{tabular} & \begin{tabular}{@{}c@{}}Eden ($M_{e}$)\end{tabular} \\
 \hline
 \end{tabular}
}%
\vspace*{-6mm}
\label{tab:options}
\end{table}

Allocation to the pools \poolcache\ ($M_{c}$) and \poolshuffle\ ($M_{s}$) is controlled by application frameworks both to make an efficient use of available memory
 and in order to avoid {\em out-of-memory} errors. 
Spark, for example, provides a configuration option called {\em spark.memory.fraction} to bound the two pools~\cite{spark-docs}; Configuration {\em taskmanager.memory.fraction} plays a similar role in Flink~\cite{flink-docs}. The other two memory pools \poolcode\ ($M_{i}$) and \poolunmgd\ ($M_{u}$), however, are not managed explicitly.

Data analytics applications include multiple {\em stages} of computations where the stages are divided by shuffle dependencies. Computations within a stage are parallelized into a number of tasks each processing one partition of the input data. Although all tasks from a stage can be run in parallel, they are scheduled in multiple {\em waves} of execution. The number of tasks in a wave is determined by the number of running containers and the number of execution {\em slots} available on each container. (The number of tasks in a wave is similar to the concept of MPL in parallel database systems~\cite{mpl}.) The number of slots on a container is a configuration parameter which is determined based on the amount of resources (CPU, memory, I/O) expected to be consumed by a task. 

In summary, Table~\ref{tab:options} lists the parameters controlling usage of memory pools in---and effectively impacting the performance of---memory-based analytics systems.

\vspace*{-4mm}
\subsection{Application Tuning}
\label{sec:tuning}
\vspace*{-1mm}

Users of data analytics systems expect to achieve the best possible latency (wall clock duration) for their applications. In memory-based analytics, the performance is largely dependent on the {\em safety} and {\em efficiency} of the memory usage. Under this framework, an application can be tuned at the following levels: 
(a) while allocating resources from the resource managers, (b) while setting options provided by the application framework related to the degree of parallelism or internal memory pools among others, and (c) while configuring JVM parameters related to garbage collection of heap. 
Applications we consider for tuning constitute a given workflow (or query plan) and a given input data. Re-using tuning results when any of these inputs changes is left out of the scope of this paper.
We first outline three broad categories of tuning approaches possible for our problem setup before describing our solution.

\noindent{\bf I. Robust defaults:}
Cloud vendors and application frameworks provide default settings for certain parameters that are expected to generalize towards a broad spectrum of applications. Amazon's popular cloud-based offering Elastic MapReduce (EMR) provides a default policy for resource allocation on Spark clusters, called \maximizeresource~\cite{spark-emr}. This policy creates a single resource container on each worker node allocating it the entire compute and memory resources. The expectation is that the containers will perform the best when allocated the maximum possible resources. Application frameworks such as Spark and Flink provide default settings for application level and JVM level memory pools~\cite{spark-docs, flink-docs}. The defaults use heuristics that generalize well, e.g., \poolold\ pool size is set higher than the value chosen for \poolcache\ so that the long living cache objects can fit in the tenured space. However, the defaults leave a lot of scope for performance improvements which can be  exploited easily by expert users~\cite{ryza, thoth-action} on a per-application basis. 

\noindent{\bf II. White-box modeling:}
One approach towards building an automated tuning solution is to build a thorough understanding of the impact of configuration options on application performance and use it towards developing analytical {\em What-If} models for performance estimations. Solutions exist in DBMSes~\cite{db2advisor}, or in MapReduce systems~\cite{starfish, mrtuner}. But developing such models is nontrivial ~\cite{wishful} or downright impossible given the wide variety in the computational models and the physical design of data to consider. Most of the literature has focussed on training ML-based performance models using either a small-scale benchmark test bed, historical performance data, or from application performance under low workload~\cite{ernest, ottertune, cart, spark-trial, spark-tune-1, cdbtune, deeprm}. However, the understanding developed by these {\em offline} approaches may not directly help tune a new application, or may potentially involve a long {\em online} learning cycle. 

\noindent{\bf III. Black-box modeling:}
Black-box approaches are often employed when building an understanding of the interactions among configuration options either analytically or through offline training is impractical. Search-based tuning approaches to find the optimal configuration typically involve a combination of random sampling and local search~\cite{smarthill, mronline, elastisizer, bdaf, bestconfig}. However, such approaches can result in a very expensive exploration given the complex non-linear response surfaces and the high costs associated with running each experiment. A better option is to employ an improvement-based policy which follows a Sequential Model-based Optimization (SMBO) approach~\cite{smac}. SMBO iterates between fitting a surrogate model and using it to recommend the next probe of the configuration space. Bayesian optimization (\bayesian)~\cite{bayesian-book} is a powerful state-of-the-art SMBO technique that is applied to varied designs including Database systems~\cite{ituned, ottertune}, Streaming~\cite{bo4co}, Storage systems~\cite{storage-tuning}, and Cloud infrastructures~\cite{cherrypick, arrow}. We consider \bayesian\ as a candidate black-box policy for our problem domain as it provides a theoretically-justified way to explore the configuration space with improvement guarantees. 

Another popular AI-based policy we consider is Deep Deterministic Policy Gradient (\reinforce)~\cite{ddpg}. It provides a powerful reinforcement learning algorithm that is hugely popular in the fields of robotics and imaging and has recently been adopted in database systems~\cite{qtune, cdbtune}. \reinforce\ combines Deep Q Network with Actor-Critic models to automatically learn the best policy to quickly reach to the most optimal state which corresponds to the configuration of the memory parameters in our problem domain.

Our evaluation shows that despite the advances in the black-box tuning approaches, the number of experiments (test runs) required to have sufficient confidence in predictions could still be significant. 
This number could be lowered if we could use some internal understanding of the impact of the memory configurations. We carry out an empirical study (presented in Section~\ref{sec:interactions}) to develop a deep understanding of the various interactions among the configuration options and the resource usage metrics. The empirical study is used in building an analytical algorithm, called \relm, to recommend a configuration which ensures both a reliable as well as a resource-efficient execution of the application. \relm\ relies on a single application profile to learn application-specific requirements of the various resources required for different processing needs. The requirements obtained from the profile are fed to a set of analytical models which combine, in quick time, to recommend a configuration most suited to the application's needs. Details of the design of \relm\ are provided in Section~\ref{sec:tuner}.

In another important contribution, the analytical models developed in \relm\ are used to speed-up the exploration process in \bayesian. The modification, called Guided Bayesian Optimization (\guided), is detailed in Section~\ref{sec:guided}. The idea behind \guided\ is to plug-in the system internal knowledge in the form of a small number of simple analytical models as extra parameters to the surrogate model of \bayesian. These extra parameters help the model learn the distinction between the expensive (undesired) regions of the configurations and the inexpensive (desired) regions in quick time.


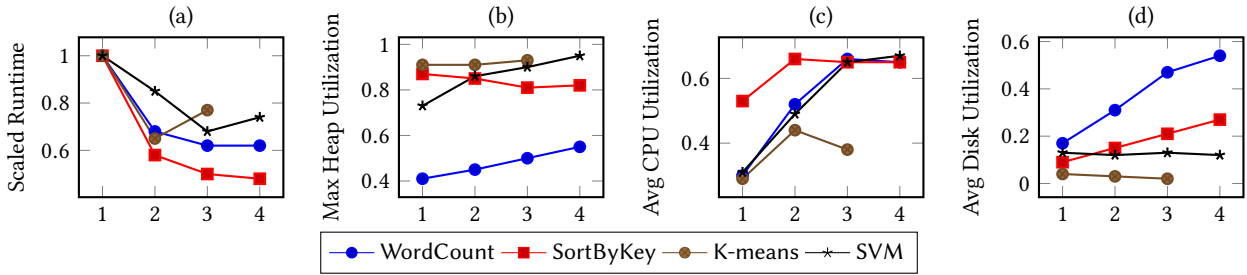
\begin{figure*}
\centering
\begin{tikzpicture}
\begin{axis}[
title={(a)},
title style={yshift=-1.5ex},
ylabel={Scaled Runtime},
legend columns=-1,
legend entries={\wc, \sort, \kmeans, \svm, \pagerank},
legend to name=named,
enlargelimits=0.15,
smooth, thick,
]
\addplot+[sharp plot] coordinates
{(1,1) (2,.68) (3,.62) (4,.62)};   
\addplot+[sharp plot] coordinates
{(1,1) (2,.58) (3,.5) (4,.48)};
\addplot+[sharp plot] coordinates
{(1,1) (2,.65) (3,.77) 
};    
\addplot+[sharp plot] coordinates
{(1,1) (2,.85) (3,.68) (4,.74)};    
\addplot+[sharp plot] coordinates
{};
\end{axis}
\end{tikzpicture}
\quad
\begin{tikzpicture}
\begin{axis}[
title={(b)},
title style={yshift=-1.5ex},
ylabel={Max Heap Utilization},
enlargelimits=0.15,
smooth, thick,
]
\addplot+[sharp plot] coordinates
{(1,.41) (2,.45) (3,.5) (4,.55)};    
\addplot+[sharp plot] coordinates
{(1,.87) (2,.85) (3,.81) (4,.82)};    
\addplot+[sharp plot] coordinates
{(1,.91) (2,.91) (3,.93) 
};    
\addplot+[sharp plot] coordinates
{(1,.73) (2,.86) (3,.9) (4,.95)};    
\end{axis}
\end{tikzpicture}
\quad
\begin{tikzpicture}
\begin{axis}[
title={(c)},
title style={yshift=-1.5ex},
ylabel={Avg CPU Utilization},
enlargelimits=0.15,
smooth, thick,
]
\addplot+[sharp plot] coordinates
{(1,.3) (2,.52) (3,.66) (4,.65)};    
\addplot+[sharp plot] coordinates
{(1,.53) (2,.66) (3,.65) (4,.65)};    
\addplot+[sharp plot] coordinates
{(1,.29) (2,.44) (3,.38) 
};    
\addplot+[sharp plot] coordinates
{(1,.31) (2,.49) (3,.65) (4,.67)};    
\end{axis}
\end{tikzpicture}
\quad
\begin{tikzpicture}
\begin{axis}[
title={(d)},
title style={yshift=-1.5ex},
ylabel={Avg Disk Utilization},
enlargelimits=0.15,
smooth, thick,
]
\addplot+[sharp plot] coordinates
{(1,.17) (2,.31) (3,.47) (4,.54)};    
\addplot+[sharp plot] coordinates
{(1,.09) (2,.15) (3,.21) (4,.27)}; 
\addplot+[sharp plot] coordinates
{(1,.04) (2,.03) (3,.02) 
};    
\addplot+[sharp plot] coordinates
{(1,.13) (2,.12) (3,.13) (4,.12)};    
\end{axis}
\end{tikzpicture}
\\
\ref{named}
\vspace*{-4mm}
\caption{Impact of increasing number of containers per node on runtime (a), maximum heap utilization (b), average CPU utilization (c), and average disk utilization (d) on benchmark applications. Missing points correspond to instances of failures.}
\vspace*{-6mm}
\label{fig:execs}
\end{figure*}

\begin{figure}
\centering
\begin{tikzpicture}
\begin{axis}[
width=5.5cm,
height=4.5cm,
ylabel={Runtime (mins)},
legend columns=1,
legend entries={\sort, \kmeans, \pagerank},
xticklabels={},
legend pos=outer north east,
enlargelimits=0.1,
smooth, thick,
]
\addplot+[scatter, 
only marks, 
scatter/classes={a={}},
scatter src=explicit symbolic, 
nodes near coords*={\label},
visualization depends on={value \thisrow{label} \as \label}
] table [meta=class] 
{
 x y class label
}; 
\addplot+[scatter, 
only marks, 
scatter/classes={a={}},
scatter src=explicit symbolic, 
nodes near coords*={\label},
visualization depends on={value \thisrow{label} \as \label}
] table [meta=class] 
{
 x y class label
 1 28 a $4^*$
 2 16 a $5^*$
 3 44 a 4
 4 31 a $5^*$
 5 40 a 1
}; 
\addplot+[scatter, 
only marks, 
scatter/classes={a={}},
scatter src=explicit symbolic, 
nodes near coords*={\label},
visualization depends on={value \thisrow{label} \as \label}
] table [meta=class] 
{
 x y class label
 6 19 a 0
 7 15 a $4^*$
 8 19 a 0
 9 12 a $3^*$
 10 25 a 2
}; 
\addplot+[scatter, 
only marks, 
scatter/classes={a={}},
scatter src=explicit symbolic, 
nodes near coords*={\label},
visualization depends on={value \thisrow{label} \as \label}
] table [meta=class] 
{
 x y class label
}; 
\addplot+[scatter, 
only marks, 
scatter/classes={a={}},
scatter src=explicit symbolic, 
nodes near coords*={\label},
visualization depends on={value \thisrow{label} \as \label}
] table [meta=class] 
{
 x y class label
 11 66 a $6^*$ 
 12 54 a 1
 13 55 a 4
 14 70 a $6^*$
 15 45 a $9^*$
}; 
\end{axis}
\end{tikzpicture}
\vspace*{-6mm} 
\caption{Exploring failures on one {\em unsafe} configuration each for \sort, \kmeans, and \pagerank, namely: (1) assigning 70\% heap for shuffle, (2) running 4 containers per node, and (3) keeping default settings. Each setup is executed 5 times. Point labels indicate the number of container failures during the run, with * marks denoting aborted runs.}
\vspace*{-7mm} 
\label{fig:runs}
\end{figure}
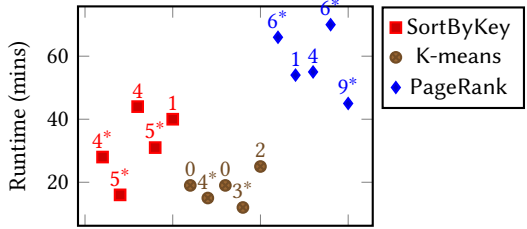

\vspace*{-2mm}
\section{Understanding interactions}
\label{sec:interactions}
\vspace*{-1mm}

Data analytics applications  vary widely  in their computational model (e.g., SQL, shuffling, iterative processing) and physical design of input data (e.g., partition sizes). This translates to variations in resource consumption patterns of the computations. We have listed the most important memory configuration options in Table~\ref{tab:options}. Here, we explore the impact of each option using the five representative benchmark applications listed in Table~\ref{tab:benchmarks}. The test suite covers a broad spectrum of computational models and physical designs making it ideal for the empirical study. All experiments were carried out on Cluster A listed in Table~\ref{tab:setup}.

\vspace*{-2mm}
\subsection{Containers per Node} \label{sec:containers}
\vspace*{-1mm}

As shown in Figure~\ref{fig:rmlevel}, physical memory on a worker node is divided into multiple containers by the resource manager. This creates a spectrum of choices for an application from using a small number of {\em fat} containers to a large number of {\em thin} containers. The default policy used on Amazon EMR clusters, called \maximizeresource, creates one fat container on each node assigning the entire node memory (minus OS overheads) to it. 
We vary the number of containers on a node from 1 to 4. The corresponding \poolheap\ allocation shrinks from 4404MB to 1101MB proportionately. The other parameters are set to their default values as listed in Table~\ref{tab:default}. 

Figure~\ref{fig:execs} shows the results. Only the successful application runs are included in the plots, \pagerank\ is entirely missing as it fails under each setting including the default setup. Failures will be discussed separately.
From the runtimes (normalized to the runtimes on the default setup), it can be noticed that \wc\ and \sort\ perform significantly better on thin containers. Both the applications do not use any cache storage and are, therefore, less memory-bound compared to the machine learning applications, namely, \kmeans\ and \svm. However, the performance does not scale linearly because of the CPU and Disk bottlenecks as indicated by an increase in the corresponding resource utilization metrics. Tasks running \kmeans\ and \svm\ get less memory for processing because of cache storage. As a result, thin containers run into memory pressures leading to a degradation of performance. \kmeans, in fact, runs into {\em out-of-memory} failures with 4 containers per node. 
This analysis shows how the application task memory requirements play an important role in tuning the executors.

\noindent \textit{\textbf{Observation 1:} Containers should be adequately sized to just meet the cache and the task memory requirements.}

\noindent {\bf Failure cases.} 
Results presented in Figure~\ref{fig:execs} do not include \pagerank\ application because it fails under each setup. 
We probe three setups next, one each for \sort, \kmeans\ and \pagerank, where containers were observed to fail. Each setup is executed 5 times. Figure~\ref{fig:runs} shows the results. 
Failures observed here are caused by two reasons: (a) {\em Out-of-memory} errors while creating objects on heap for either input data deserialization or as buffers fetching data over the network; (b) Resource manager {\em killing} containers that exceed a preset limit for physical memory usage. A container failure does not necessarily translate to application failure. Spark, our evaluation system, requests new containers to replace the failed ones and retries the failed tasks. If a task fails a pre-specified number of times, the entire application job is aborted. We have marked such occurrences separately in the graph. It can be noticed that there is a huge variability both in terms of the number of container failures and the possibility of application failure under each setup. The runtime of each application, too, is highly unpredictable. 

\noindent \textit{\textbf{Observation 2:} Over-provisioning for internal memory pools can result in unreliable performance.}

\begin{table}[!t]
\centering
\caption{Test suite used in evaluation}
\vspace*{-4mm} 
\resizebox{\columnwidth}{!}{%
 \begin{tabular}{|c c c c|} 
 \hline
 Application & Category & Dataset & Partition Size \\
 \hline
 \wc & Map and Reduce & \begin{tabular}{@{}c@{}}Hadoop RandomTextWriter (50GB)\end{tabular} & 128MB \\
 \sort & Map and Reduce & \begin{tabular}{@{}c@{}}Hadoop RandomTextWriter (30GB)\end{tabular} & 512MB \\
 \kmeans & Machine Learning & \begin{tabular}{@{}c@{}}HiBench huge (100M samples)\end{tabular} & 128MB \\
 \svm & Machine Learning & \begin{tabular}{@{}c@{}}HiBench huge (100M examples)\end{tabular} & 32MB \\
 \pagerank & Graph & \begin{tabular}{@{}c@{}}LiveJournal~\cite{livejournal} (69M edges)\end{tabular} & 128MB \\
 TPC-H & SQL & \begin{tabular}{@{}c@{}}TPC-H DBGen (50 scale factor)\end{tabular} & 128MB \\
 \hline
 \end{tabular}
}%
\label{tab:benchmarks}
\caption{Evaluation cluster setups}
\vspace*{-4mm} 
\resizebox{.68\columnwidth}{!}{
 \begin{tabular}{| l | c | c |} 
 \hline
 & Cluster A & Cluster B \\
 \hline
 Node types & Physical & Virtual EC2 \\
 Number of nodes & 8 & 4 \\
 Memory per node  & 6GB & 32GB \\
 CPU cores per node & 8 & 31 ECU \\
 Network bandwidth & 1Gbps & 10Gbps \\
 \hline
 Compute Framework & \multicolumn{2}{c|} {Spark-2.0.1} \\
 Resource Manager & \multicolumn{2}{c|} {Yarn-2.7.2} \\
 JVM Framework & \multicolumn{2}{c|} {OpenJDK-1.8.0} \\
 \hline
 \end{tabular}
}
\label{tab:setup}
\caption{Config values suggested by \maximizeresource\ and framework defaults on Cluster A.}
\vspace*{-4mm} 
\resizebox{.6\columnwidth}{!}{
 \begin{tabular}{|l | l|} 
 \hline
 \begin{tabular}{@{}c@{}}Containers per Node\end{tabular} & 1 \\ 
 \begin{tabular}{@{}c@{}}Heap Size\end{tabular} & 4404MB \\ 
 \begin{tabular}{@{}c@{}}Task Concurrency\end{tabular} & 2 \\ 
 \begin{tabular}{@{}c@{}}Cache Capacity + Shuffle Capacity\end{tabular} & .6 \\
 NewRatio & 2 \\
 SurvivorRatio & 8\\
 \hline
 \end{tabular}
}
\label{tab:default}
\vspace*{-6mm} 
\end{table}


\vspace*{-3mm}
\subsection{Task Concurrency}
\label{sec:concurrency}
\vspace*{-1mm}

\begin{figure*}
\centering
\begin{tikzpicture}
\begin{axis}[
title={(a)},
ylabel={Scaled Runtime},
title style={yshift=-1.5ex},
legend columns=-1,
legend entries={\wc, \sort, \kmeans, \svm, \pagerank},
legend to name=named,
enlargelimits=0.15,
smooth, thick,
]
\addplot+[sharp plot] coordinates
{(1,1) (2,.65) (3,.55) (4,.52) (5,.52) (6,.52) (7,.54) (8,.54)};    
\addplot+[sharp plot] coordinates
{(1,1) (2,.91) (3,.80) (4,.67) (5,.65) (6,.63) (7,.61) (8,.61)};
\addplot+[sharp plot] coordinates
{(1,1) (2,.81) (3,.63) (4,.62) (5,.59) (6,.65) (7,.62) (8,.62)};    
\addplot+[sharp plot] coordinates
{(1,1) (2,.65) (3,.64) (4,.58) (5,.56) (6,.57) (7,.63) (8,.65)};    
\addplot+[sharp plot] coordinates
{(1,1)
 };    
\end{axis}
\end{tikzpicture}
\quad
\begin{tikzpicture}
\begin{axis}[
title={(b)},
title style={yshift=-1.5ex},
ylabel={Max Heap Utilization},
enlargelimits=0.15,
smooth, thick,
]
\addplot+[sharp plot] coordinates
{(1,.4) (2,.4) (3,.4) (4,.41) (5,.42) (6,.42) (7,.42) (8,.42)};    
\addplot+[sharp plot] coordinates
{(1,.79) (2,.86) (3,.87) (4,.88) (5,.88) (6,.88) (7,.87) (8,.87)};    
\addplot+[sharp plot] coordinates
{(1,.87) (2,.89) (3,.9) (4,.91) (5,.93) (6,.92) (7,.93) (8,.92)};    
\addplot+[sharp plot] coordinates
{(1,.73) (2,.74) (3,.77) (4,.82) (5,.86) (6,.87) (7,.88) (8,.88)};    
\addplot+[sharp plot] coordinates
{(1,.93) 
};    
\end{axis}
\end{tikzpicture}
\quad
\begin{tikzpicture}
\begin{axis}[
title={(c)},
title style={yshift=-1.5ex},
ylabel={Avg CPU Utilization},
enlargelimits=0.15,
]
\addplot+[sharp plot] coordinates
{(1,.18) (2,.31) (3,.44) (4,.59) (5,.63) (6,.67) (7,.71) (8,.70)};    
\addplot+[sharp plot] coordinates
{(1,.25) (2,.48) (3,.57) (4,.54) (5,.55) (6,.54) (7,.54) (8,.53)};    
\addplot+[sharp plot] coordinates
{(1,.20) (2,.29) (3,.36) (4,.41) (5,.43) (6,.43) (7,.44) (8,.45)};    
\addplot+[sharp plot] coordinates
{(1,.18) (2,.29) (3,.34) (4,.42) (5,.50) (6,.55) (7,.52) (8,.53)};    
\addplot+[sharp plot] coordinates
{(1,.29) 
};    
\end{axis}
\end{tikzpicture}
\quad
\begin{tikzpicture}
\begin{axis}[
title={(d)},
title style={yshift=-1.5ex},
ylabel={Avg Disk Utilization},
enlargelimits=0.15,
smooth, thick,
]
\addplot+[sharp plot] coordinates
{(1,.15) (2,.20) (3,.45) (4,.62) (5,.60) (6,.60) (7,.66) (8,.66)};    
\addplot+[sharp plot] coordinates
{(1,.05) (2,.07) (3,.10) (4,.13) (5,.13) (6,.13) (7,.15) (8,.14)};    
\addplot+[sharp plot] coordinates
{(1,.03) (2,.01) (3,.02) (4,.01) (5,.03) (6,.02) (7,.01) (8,.03)};    
\addplot+[sharp plot] coordinates
{(1,.12) (2,.11) (3,.11) (4,.11) (5,.13) (6,.15) (7,.12) (8,.11)};    
\addplot+[sharp plot] coordinates
{(1,.02) 
};    
\end{axis}
\end{tikzpicture}
\\
\ref{named}
\vspace*{-4mm}
\caption{Impact of Task Concurrency on runtime (a), maximum heap utilization (b), average CPU utilization (c), and average disk utilization (d) for benchmark applications. \pagerank\ runs out of memory for Task Concurrency$\geq 2$.}
\vspace*{-5mm}
 \label{fig:dop}
\end{figure*}
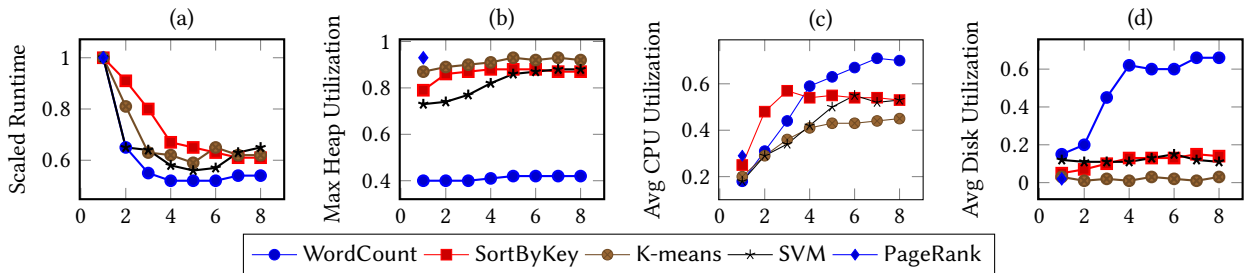

An important optimization to increase throughput of the application tasks is to increase concurrency. We study this optimization in Figure~\ref{fig:dop} changing Task Concurrency from 1 to 8. The runtimes are normalized to the setup with task concurrency set to 1. The performance of each application improves with concurrency until a certain point beyond which it plateaus out. 

We include Maximum Heap Utilization,  Average CPU Utilization, and Average Disk Utilization plots to inspect the possible bottlenecks causing the plateau effect. For all applications except \wc, the effect can be explained by memory pressures indicated by the maximum heap utilization numbers. As all concurrently running tasks in a container have to compete for a fixed sized heap, increasing task concurrency leads to more garbage collection overheads, curtailing the benefits of the increased parallelism. Tasks for \wc, though not bottlenecked by memory, suffer from CPU and disk bottlenecks on higher values of concurrency. 

\noindent \textit{\textbf{Observation 3:} Resource bottlenecks including CPU, I/O, and memory must be accounted for while increasing Task Concurrency.}

\vspace*{-3mm}
\subsection{Cache and Shuffle memory}\label{sec:cache}
\vspace*{-1mm}

\begin{figure*}
\centering
\begin{tikzpicture}
\begin{axis}[
title={(a)},
title style={yshift=-1.5ex},
ylabel={Scaled Runtime},
legend columns=-1,
legend entries={\wc, \sort, \kmeans, \svm, \pagerank},
legend to name=named,
enlargelimits=0.15,
smooth, thick,
]
\addplot+[sharp plot] coordinates
{(.1,1) (.2,.92) (.3,.92) (.4,.92) (.5,.94) (.6,.92) (.7,.94) (.8,.94) (.9,.97)};
\addplot+[sharp plot] coordinates
{(.1,1) (.2,1.14) (.3,1.27) (.4,1.4) (.5,1.77) (.6,2)  
};    
\addplot+[sharp plot] coordinates
{(.1,1) (.2,.88) (.3,.79) (.4,.69) (.5,.62) (.6,.60) (.7,.57)  
};    
\addplot+[sharp plot] coordinates
{(.1,1) (.2,.92) (.3,.85) (.4,.73) (.5,.69) (.6,.57) (.7,.57) (.8,.58) (.9,.69)};    
\addplot+[sharp plot] coordinates
{(.1,1) (.2,1) (.3,1) (.4,.94) (.5,.94) (.6,.89) (.7,.91) 
};    
\end{axis}
\end{tikzpicture}
\quad
\begin{tikzpicture}
\begin{axis}[
title={(b)},
title style={yshift=-1.5ex},
ylabel={Max Heap Utilization},
enlargelimits=0.15,
smooth, thick,
]
\addplot+[sharp plot] coordinates
{(.1,.4) (.2,.4) (.3,.4) (.4,.4) (.5,.4) (.6,.4) (.7,.4) (.8,.4) (.9,.39)};    
\addplot+[sharp plot] coordinates
{(.1,.61) (.2,.76) (.3,.78) (.4,.79) (.5,.83) (.6,.85)};    
\addplot+[sharp plot] coordinates
{(.1,.83) (.2,.9) (.3,.93) (.4,.92) (.5,.93) (.6,.89) (.7,.88)};    
\addplot+[sharp plot] coordinates
{(.1,.49) (.2,.57) (.3,.65) (.4,.71) (.5,.75) (.6,.74) (.7,.74) (.8,.74) (.9,.74)};    
\addplot+[sharp plot] coordinates
{(.1,.93) (.2,.91) (.3,.92) (.4,.93) (.5,.92) (.6,.92) (.7,.91)
};    
\end{axis}
\end{tikzpicture}
\quad
\begin{tikzpicture}
\begin{axis}[
title={(c)},
title style={yshift=-1.5ex},
ylabel={GC Overheads},
enlargelimits=0.15,
smooth, thick,
]
\addplot+[sharp plot] coordinates
{(.1,.08) (.2,.09) (.3,.09) (.4,.09) (.5,.09) (.6,.09) (.7,.09) (.8,.09) (.9,.1)};    
\addplot+[sharp plot] coordinates
{(.1,.18) (.2,.28) (.3,.36) (.4,.43) (.5,.48) (.6,.59)};    
\addplot+[sharp plot] coordinates
{(.1,.11) (.2,.12) (.3,.14) (.4,.16) (.5,.18) (.6,.22) (.7,.3)};    
\addplot+[sharp plot] coordinates
{(.1,.1) (.2,.1) (.3,.1) (.4,.1) (.5,.1) (.6,.1) (.7,.1) (.8,.1) (.9,.1)};    
\addplot+[sharp plot] coordinates
{(.1,.17) (.2,.17) (.3,.18) (.4,.17) (.5,.18) (.6,.18) (.7,.22) 
};    
\end{axis}
\end{tikzpicture}
\quad
\begin{tikzpicture}
\begin{axis}[
title={(d)},
title style={yshift=-1.5ex},
ylabel={Cache Hit Ratio},
enlargelimits=0.15,
smooth, thick,
]
\addplot+[sharp plot] coordinates
{};    
\addplot+[sharp plot] coordinates
{};    
\addplot+[sharp plot] coordinates
{(.1,.06) (.2,.18) (.3,.31) (.4,.44) (.5,.57) (.6,.71) (.7,.85)};    
\addplot+[sharp plot] coordinates
{(.1,.21) (.2,.42) (.3,.62) (.4,.83) (.5,.98) (.6,1) (.7,1) (.8,1) (.9,1)};    
\addplot+[sharp plot] coordinates
{(.1,.03) (.2,.09) (.3,.17) (.4,.25) (.5,.32) (.6,.41) (.7,.48) 
};    
\end{axis}
\end{tikzpicture}
\\
\ref{named}
\vspace*{-4mm}
\caption{Impact of Cache Capacity and Shuffle Capacity on runtime (a), maximum heap utilization (b), and average per task GC Overheads (c) for benchmark applications. The X-axis represents Shuffle Capacity for \wc\ and \sort\ which do not use any cache. In case of applications \kmeans, \svm, and \pagerank\ which predominantly use cache; the X-axis represents Cache Capacity as a fraction of allocated heap. The cache hit ratio for these applications is displayed in plot (d).}
\vspace*{-5mm}
 \label{fig:fraction}
\end{figure*}
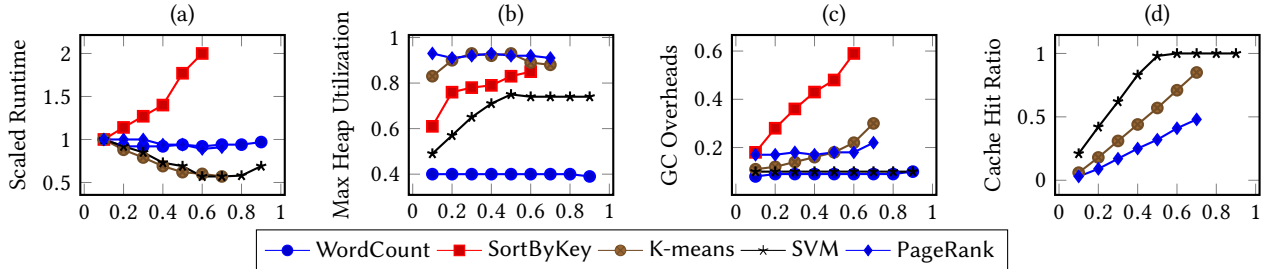

We explore impact of the memory allocated to internal memory pools of \poolcache\ and \poolshuffle\ on our benchmark applications; the results are included in Figure~\ref{fig:fraction}. Since Spark uses a unified memory pool~\cite{unified} to manage both, we vary a single parameter that changes the fraction of heap allocated to the unified pool. Further, we single out the applications \kmeans, \svm, and \pagerank\ for the analysis of Cache Capacity as they predominantly use cache. Applications \wc\ and \sort\, on the other hand, are analyzed for the shuffle memory since they use the unified memory pool exclusively for shuffle objects. Task concurrency for \pagerank\ is set to 1 while the other applications use the default setting of 2: This change is done in order to avoid {\em out-of-memory} errors on higher Task Concurrency values observed for \pagerank.

It can be noticed that an increase in Cache Capacity results in performance gains for each of the \kmeans, \svm, and \pagerank\ applications until a certain value beyond which either the performance plateaus or containers run out of memory. We include a plot showing `Cache Hit Ratio' (Figure~\ref{fig:fraction}(d)) which gives a ratio of the number of data partitions found in cache over the total number of partitions requested to be cached. It can be noticed that \svm\ can fit 100\% partitions in cache with a capacity over 0.5, the point where its performance plateaus. \kmeans\ hits the memory bottleneck before it can fit all the partitions. GC overheads, derived by averaging the fraction of time spent by tasks in GC processes, also indicate a sharp rise before the containers fail at a Cache Capacity of 0.8. The same effect is seen for \pagerank\ as well. 

\noindent \textit{\textbf{Observation 4:} Leave sufficient memory for tasks while optimizing for cache storage.}

Analysis of shuffle memory throws the most counter-intuitive result for \sort\ where assigning more shuffle memory leads to performance degradation. Tasks running the reduce stage of \sort\ use memory to perform an in-memory sort of data. If the memory allocated from shuffle pool is insufficient, then the tasks use an external merge-sort by spilling partially sorted records to disk and merging them later. Although increasing shuffle memory leads to lowering the number of spills, increased size of each spill puts more pressure on garbage collection. The GC overheads plot shows that tasks spend 60\% time on average in garbage collections for a Shuffle Capacity of 0.6. We analyze an interplay between the shuffle pool size and GC settings in the next subsection which clearly explains why higher Shuffle Capacity settings are undesirable. 

\vspace*{-3mm}
\subsection{Interactions with GC settings}
\label{sec:gc}
\vspace*{-1mm}

\begin{figure*}
\centering
\begin{tikzpicture}
\begin{axis}[
title={Runtime (min)},
ylabel={NewRatio},
title style={yshift=-1.5ex},
view={0}{90},
xtick=data, ytick=data,
colorbar,
y dir=reverse,
enlargelimits=0.1,
]
\addplot3[surf, shader=faceted interp, scatter, mark=*, mesh/rows=5] coordinates
{
 (0.4, 1, 27) (0.4, 2, 26) (0.4, 3, 27) (0.4, 4, 28)
 (0.5, 1, 25) (0.5, 2, 23) (0.5, 3, 25) (0.5, 4, 25)
 (0.6, 1, 25) (0.6, 2, 22) (0.6, 3, 22) (0.6, 4, 21)
 (0.7, 1, 27) (0.7, 2, 21) (0.7, 3, 20) (0.7, 4, 18)
 (0.8, 1, 41) (0.8, 2, 43) (0.8, 3, 21) (0.8, 4, 16)
};
\end{axis}
\end{tikzpicture}
\quad
\begin{tikzpicture}
\begin{axis}[
title={GC Overheads},
ylabel={NewRatio},
title style={yshift=-1.5ex},
view={0}{90},
xtick=data, ytick=data,
colorbar,
y dir=reverse,
enlargelimits=0.1,
]
\addplot3[surf, shader=faceted interp, scatter, mark=*, mesh/rows=5] coordinates
{
 (0.4, 1, .14) (0.4, 2, .14) (0.4, 3, .16) (0.4, 4, .15)
 (0.5, 1, .19) (0.5, 2, .18) (0.5, 3, .19) (0.5, 4, .18)
 (0.6, 1, .26) (0.6, 2, .2) (0.6, 3, .22) (0.6, 4, .2)
 (0.7, 1, .45) (0.7, 2, .27) (0.7, 3, .3) (0.7, 4, .27)
 (0.8, 1, .55) (0.8, 2, .47) (0.8, 3, .37) (0.8, 4, .32)
};
\end{axis}
\end{tikzpicture}
\quad
\begin{tikzpicture}
\begin{axis}[
title={Cache Hit Ratio},
ylabel={NewRatio},
title style={yshift=-1.5ex},
view={0}{90},
xtick=data, ytick=data,
colorbar,
y dir=reverse,
enlargelimits=0.1,
]
\addplot3[surf, shader=faceted interp, scatter, mark=*, mesh/rows=5] coordinates
{
 (0.4, 1, .42) (0.4, 2, .44) (0.4, 3, .46) (0.4, 4, .47)
 (0.5, 1, .54) (0.5, 2, .58) (0.5, 3, .6) (0.5, 4, .63)
 (0.6, 1, .66) (0.6, 2, .72) (0.6, 3, .74) (0.6, 4, .75)
 (0.7, 1, .77) (0.7, 2, .86) (0.7, 3, .86) (0.7, 4, .88)
 (0.8, 1, .87) (0.8, 2, .83) (0.8, 3, .92) (0.8, 4, .94)
};
\end{axis}
\end{tikzpicture}
\vspace*{-0.5cm}
\caption{Impact of NewRatio and Cache Capacity on \kmeans}
\vspace*{-0.4cm}
 \label{fig:newratio}
\end{figure*}
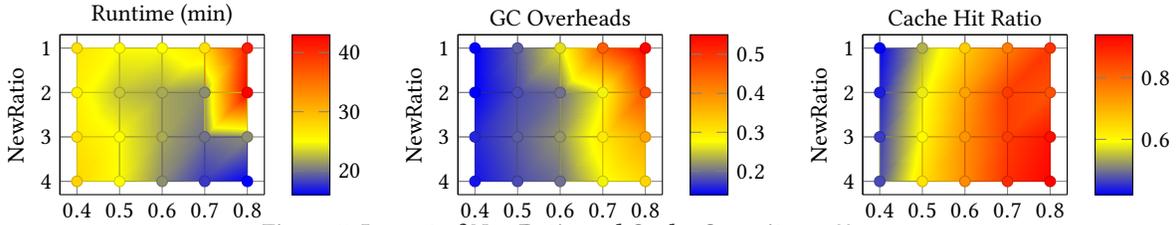

\begin{figure*}
\noindent\begin{minipage}[b]{.28\textwidth}
\vspace{0pt}
\begin{tikzpicture}
\begin{axis}[
width=\linewidth,
height=0.7\linewidth,
ylabel={GC Overheads},
smooth, thick,
enlargelimits=0.15,
]
\addplot+[sharp plot, error bars/.cd, y dir=both, y explicit] coordinates
{
 (1, .26) += (0, .16) -= (0, .16) 
 (2, .199) += (0, .15) -= (0, .15) 
 (3, .22) += (0, .16) -= (0, .16) 
 (4, .23) += (0, .16) -= (0, .16) 
 (5, .235) += (0, .162) -= (0, .162) 
 (6, .242) += (0, .16) -= (0, .16) 
 (7, .246) += (0, .157) -= (0, .157) 
 (8, .254) += (0, .163) -= (0, .163) 
};
\end{axis}
\end{tikzpicture}
\vspace*{-0.5cm}
\caption{Impact of NewRatio on per task GC Overheads for \kmeans\ with a Cache Capacity of 0.6. Error bars indicate standard deviation.}
 \label{fig:gc}
\end{minipage}%
\hfill
\begin{minipage}[b]{0.64\textwidth}
\vspace{0pt}
\centering
\begin{tikzpicture}
\begin{axis}[
width=0.4\linewidth, 
title={(a) Runtime (min)},
title style={yshift=-1.5ex},
ylabel={NewRatio},
view={0}{90},
xtick={0.1, 0.2, 0.3}, ytick=data,
colorbar,
colormap/hot2,
y dir=reverse,
enlargelimits=0.1,
]
\addplot3[surf, shader=faceted interp, scatter, mark=*, mesh/rows=6] coordinates
{
 (0.05, 1, 23) (0.05, 2, 24) (0.05, 3, 23)
 (0.10, 1, 23) (0.10, 2, 22) (0.10, 3, 29)
 (0.15, 1, 22) (0.15, 2, 26) (0.15, 3, 26)
 (0.20, 1, 27) (0.20, 2, 26) (0.20, 3, 28)
 (0.25, 1, 28) (0.25, 2, 28) (0.25, 3, 28)
 (0.30, 1, 29) (0.30, 2, 30) (0.30, 3, 31)
};
\end{axis}
\end{tikzpicture}
\begin{tikzpicture}
\begin{axis}[
width=0.4\linewidth, 
title={(b) GC Overheads},
title style={yshift=-1.5ex},
view={0}{90},
xtick={0.1, 0.2, 0.3}, ytick=data,
colorbar,
colormap/hot2,
y dir=reverse,
enlargelimits=0.1,
]
\addplot3[surf, shader=faceted interp, scatter, mark=*, mesh/rows=6] coordinates
{
 (0.05, 1, .04) (0.05, 2, .09) (0.05, 3, .16)
 (0.10, 1, .13) (0.10, 2, .26) (0.10, 3, .3)
 (0.15, 1, .17) (0.15, 2, .31) (0.15, 3, .33)
 (0.20, 1, .26) (0.20, 2, .37) (0.20, 3, .39)
 (0.25, 1, .39) (0.25, 2, .41) (0.25, 3, .43)
 (0.30, 1, .44) (0.30, 2, .48) (0.30, 3, .45)
};
\end{axis}
\end{tikzpicture}
\vspace*{-0.5cm}
\caption{Impact of NewRatio and Shuffle Capacity on runtime (a) and GC Overheads (b) for \sort}
 \label{fig:sortgc}
\end{minipage}%
\vspace*{-0.5cm}
\end{figure*}

\begin{figure}
 \centering
 \includegraphics[width=.95\columnwidth]{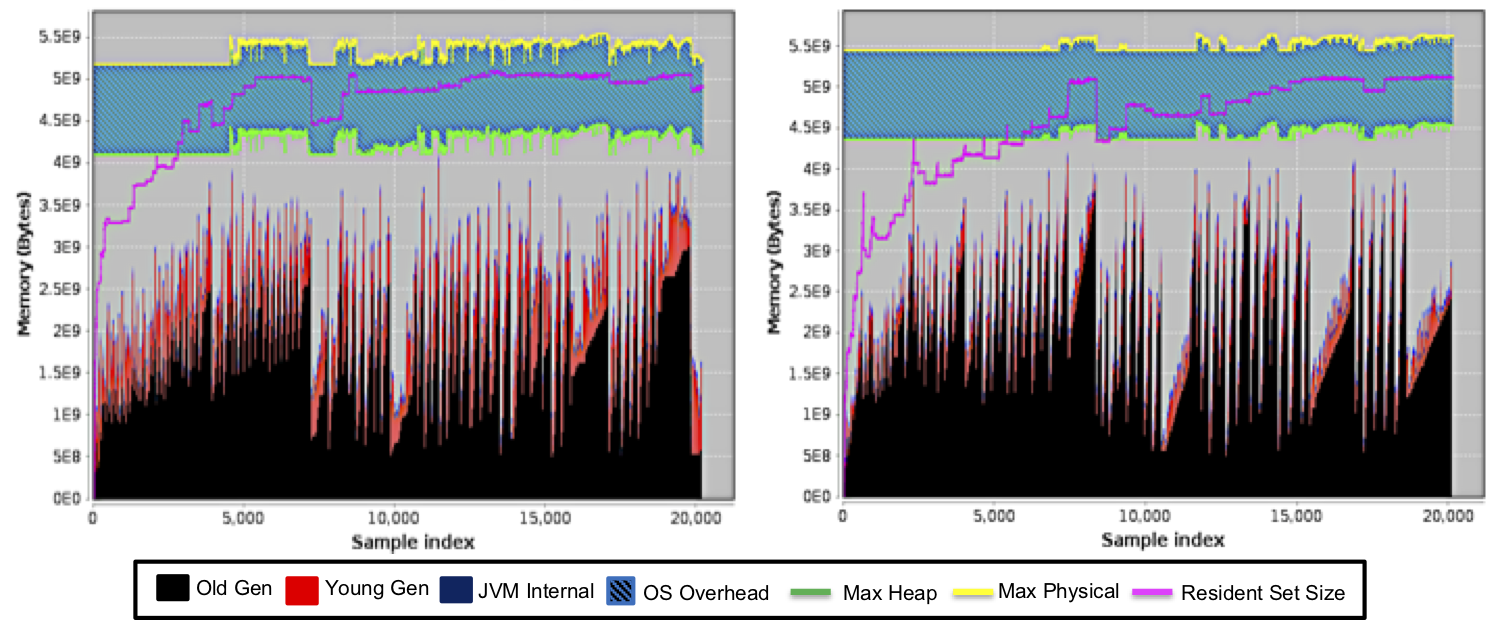}
\vspace*{-4mm} 
 \caption{Comparing memory usage timeline for a container having NewRatio=2 (left) with a container having NewRatio=5. The left side configuration is more prone to failures due to physical memory usage exceeding limit set by resource manager.}
 \vspace*{-6mm}
 \label{fig:heapusage}
\end{figure}

Section~\ref{sec:mem-overview} has described how the JVM organizes heap objects into generational pools. From an application's standpoint, this organization corresponds well to the cache requirements: As the cached objects reside in memory for a long time, they are expected to reside in the \poolold\ pool of JVM. 
We use the GC parameter NewRatio to change the capacity of \poolold\ and analyze its impact on \kmeans\ by varying the Cache Capacity from 0.4 to 0.8 in Figure~\ref{fig:newratio}. 
From the runtime numbers, the extreme results are obtained on higher Cache Capacity values (viz. 0.7 and 0.8): While the setups with lower NewRatio result in a poor performance due to large GC overheads (about 50\% of task times), the setups with higher NewRatio perform exceptionally well (3x better). It is important, therefore, to set \poolold\ pool size higher than the \poolcache\ pool. 
The key takeaway, presented below, is already known to the data engineers~\cite{spark-docs}.

\noindent \textit{\textbf{Observation 5:} Sizing \poolold\ smaller than \poolcache\ can lead to huge GC overheads (e.g., tasks spending over 50\% of their time in GC).}

The analysis above tells us to set \poolold\ size higher than \poolcache\, but how high should it be? It turns out, high values lead to increased GC overheads due to the frequent collections. Figure~\ref{fig:gc} analyzes \kmeans\ with a Cache Capacity of 0.6 with NewRatio increased from 1 to 8. Setting NewRatio to 2 provides the best outcome since it {\em just} fits the cache. Higher settings result in increasingly many invocations of {\em young GC} which add to the overheads.

The higher NewRatio settings, despite adding GC overheads, can help prevent containers exceeding physical memory usage limit set by resource managers which is one source of the container failures reported in Figure~\ref{fig:runs}. To understand this, we plot the memory usage timelines for two containers in Figure~\ref{fig:heapusage}. Lower value of NewRatio implies a lower frequency of garbage collections which results in on-heap references to the objects created in off-heap space (e.g., Native ByteBuffers used in network data transfers) getting collected less frequently. It causes the physical memory usage ({\em \textcolor{magenta} {magenta}} line showing `Resident Set Size') to grow more rapidly, and in some cases, exceeding the maximum physical memory cap ({\em \textcolor{yellow} {yellow}} line showing `Max Physical'). A higher value for NewRatio increases the frequency of garbage collection, and as a result, helps arrest the growth of physical memory.

\noindent \textit{\textbf{Observation 6:} \poolold\ capacity values larger than \poolcache\ present a trade-off between performance and reliability.}

The shuffle memory use case is very different to the cache storage. While the cached objects have a long life, the shuffle objects have a very short time span since tasks repeatedly spill the partially aggregated/ sorted results to disk multiple times during execution. Setting Shuffle Capacity larger than \pooleden\ pool size (an area within Young generation pool where newly created objects reside) necessitates a {\em full GC} every time a task spills. Figure~\ref{fig:sortgc} plots the runtimes and the GC overheads for \sort\ executed with Shuffle Capacity ranging from 0.05 to 0.3 fraction of \poolheap\ size. The NewRatio value is increased from 1 to 3 causing the \pooleden\ capacity to go down from ~37\% to ~18\% of \poolheap\ size. It should be noted that the \pooleden\ contains not only the shuffle objects but also other task objects including code data structures and partially processed data partitions. As it is hard to estimate occupancy of \pooleden\ at all times, a good heuristic could be to set the shuffle memory to 50\% of \pooleden. 

\noindent \textit{\textbf{Observation 7:} Shuffle Capacity larger than (50\% of) \pooleden\ can lead to huge GC overheads.}

\vspace*{-3mm}
\subsection{Manually tuning an application}
\label{sec:reliability}
\vspace*{-1mm}

We test our understanding by tuning \pagerank\ which 
exhibits multiple failures under the default setup (Figure~\ref{fig:runs}). The application uses {\em LiveJournalPageRank} implementation from GraphX~\cite{graphx} library on Spark. The program first coalesces input data into a small number of edge partitions. The coalesced partitions are cached in memory before running iterations on them to update page rank values of the graph nodes. Tasks running the coalesce operation need a large amount of memory 
in order to fetch partitions over the network as well as to store the partially processed partitions while more data is being {\em unrolled}. The problem is further compounded by the fact that the Cache Capacity configured for the application fits only 30\% of the cached partitions. This results in partitions being recomputed in each iteration repeating the coalesce computation.

\begin{table}
\centering
\caption{Manual tuning of \pagerank}
\vspace*{-3mm}
\resizebox{\columnwidth}{!}{%
 \begin{tabular}{|c c c c | c c c |} 
\hline
 \begin{tabular}{@{}c@{}}Containers\\ per node\end{tabular} & \begin{tabular}{@{}c@{}}Task\\ concurrency\end{tabular} & \begin{tabular}{@{}c@{}}Cache\\ Capacity\end{tabular} & \begin{tabular}{@{}c@{}}NewRatio\end{tabular} & \begin{tabular}{@{}c@{}}Runtime\\ (mins)\end{tabular} & \begin{tabular}{@{}c@{}}Cache Hit\\ Ratio\end{tabular} & \begin{tabular}{@{}c@{}}GC\\ Overheads\end{tabular} \\
 \hline
 1 & 2 & 0.6 & 2 & 66 (aborted) & 0.3 & 0.28 \\
 1 & {\bf 1} & 0.6 & 2 & 59 & 0.32 & 0.14 \\
 1 & 2 & {\bf 0.4} & 2 & 49 & 0.19 & 0.12 \\
 1 & 2 & 0.6 & {\bf 5} & 53 & 0.33 & 0.27 \\
 \hline
\end{tabular}
}%
\vspace*{-7mm}
\label{tab:fixes}
\end{table}

We try out three changes to the application configuration as listed in Table~\ref{tab:fixes}. The first row shows the default configuration. The second row lowers Task Concurrency to 1 resulting in a reliable execution (verified by running 5 times) with a runtime of 59 minutes. The third row lowers the Cache Capacity which in turn makes more memory available to tasks. This change, despite a lower cache hit ratio, provides a significant improvement to the runtime since it reduces the memory pressure. The final change we make is that of increasing NewRatio to 5 which prevents failures by collecting the physical memory used by network buffers more aggressively.

\vspace*{-3mm}
\section{RelM Tuner}
\label{sec:tuner}
\vspace*{-1mm}

The goal of \relm\ tuner is to recommend a setup of memory pools which ensures a reliable and fast performance for a data analytics application. 
In particular, \relm\ meets the following objectives:\\
{\bf (1) Safety}: Resource usage should be within allocation at all times.\\
{\bf (2a) High task concurrency}: Maximize the number of concurrently running tasks after ensuring (1).\\
{\bf (2b) High cache hit ratio}: Provision sufficient memory for cache storage after ensuring (1).\\
{\bf (3) Low GC overheads}: Limit the time spent by tasks in GC processes after ensuring (1), (2a), and (2b).

The criteria suggest a priority of goals. Safety is of foremost concern as it has the highest implications to the application performance: See Figure~\ref{fig:runs} for an example. 
We rank the goals (2a) and (2b) at the same level. Depending on application characteristics, performance is primarily a function of either one of them or both (Section~\ref{sec:interactions}). While the former is constrained by each of the CPU, memory, I/O bottlenecks, the later is constrained by the memory provisioned alone. The goal of {\em low GC overheads} is ranked the lowest in the scheme of things: Based on the settings used to meet high priority goals, we tune the parameters affecting GC overheads.

It should be noted that we do not pursue a goal of lowering shuffle data spillage here. Based on an extensive empirical study carried out by Iorgulescu et.~al.~\cite{spilled} on Hadoop, Spark, Flink, and Tez frameworks---in addition to our own evaluation presented in Section~\ref{sec:interactions}---it is evident that the memory provisioned for data shuffle has limited positive impacts on application runtime. Moreover, high values for shuffle memory could lead to GC bottlenecks as shown in Section~\ref{sec:gc}. We avoid these overheads by tuning shuffle memory and GC pool settings together as part of the goal (3).

\begin{figure}
 \centering
 \includegraphics[width=\columnwidth]{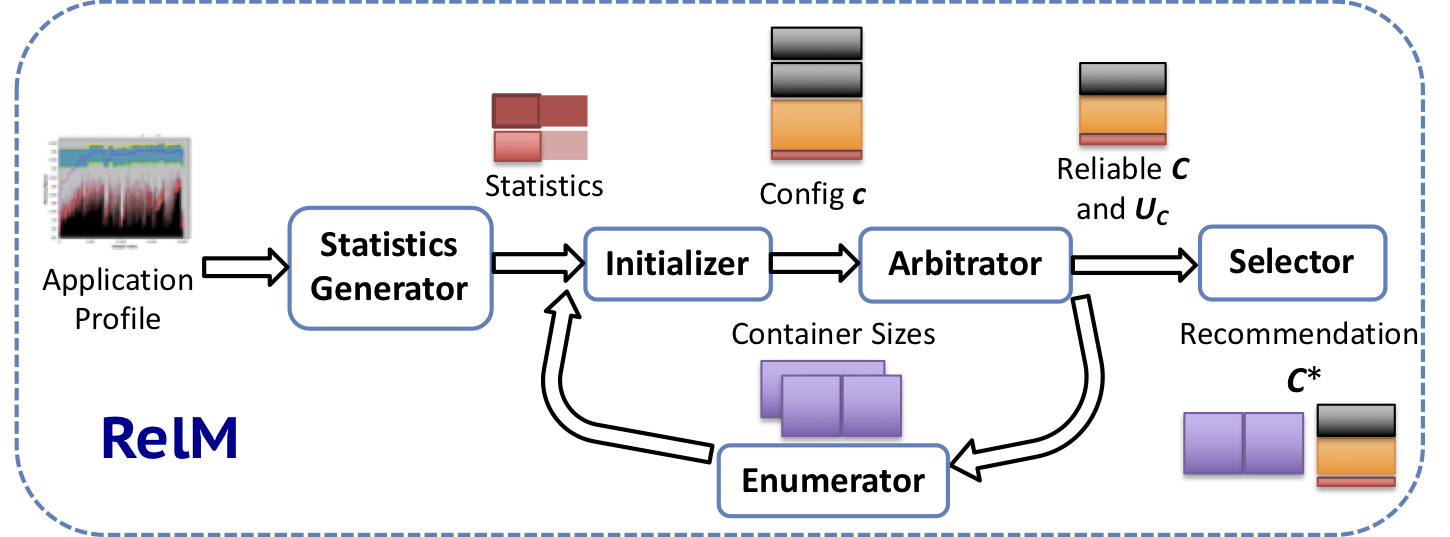}
 \vspace*{-8mm}
\caption{Application tuning process in \relm}
 \label{fig:flow}
 \vspace*{-7mm}
\end{figure}

\relm\ relies on a profile of application run to understand the resource requirements. The statistics derived from this single run are used in evaluation of all combinations of container sizes, application memory pools settings, and JVM configurations using analytical modeling. We first make a comment on the container sizes we enumerate during tuning. 
We support multiple homogeneous containers carved out of a single node with the node memory distributed equally  among them as shown in Figure~\ref{fig:rmlevel}. This gives us a small finite number of container size configurations. 

\noindent \textit{ {\bf \em Example.} Amazon EMR's m4.large nodes set the maximum memory for resource manager to 6GB with a minimum allocation size of 1GB. The possible container configurations in this case, listed as (Containers per Node, Heap Size), are: (1, 4404MB), (2, 2202MB), (3, 1468MB), and (4, 1101MB). Rest of the memory is left for OS overheads.}

\mkcomment{Does the following assumption need to be stated here?}


Figure~\ref{fig:flow} describes the steps in tuning a given application.
\squishlisttwo
 \item[1] The application profile is processed by the {\em Statistics Generator} to derive a set of statistics listed in Table~\ref{tab:stats}. (Section~\ref{sec:profiling})
 \item[2] The {\em Enumerator} module runs each container size configuration through {\em Initializer} and {\em Arbitrator}.
 \item[3] Given a container size to probe and the statistics from application profile, the {\em Initializer} module sets initial settings for memory pools optimizing for each pool independently. (Section~\ref{sec:evaluator})
 \item[4] The {\em Arbitrator} arbitrates memory assigned to various pools by the Initializer in order to ensure reliability and low GC overheads. It also calculates a utility score for the resulting configuration corresponding to its memory utilization. (Section~\ref{sec:reliable})
 \item[5] Finally, the best settings for each of the probed container size configurations are ranked by {\em Selector} based on their utility score and the best is returned as the final recommendation.
\squishend

\vspace*{-3mm}
\subsection{Statistics Generation}
\label{sec:profiling}

\begin{table}
\caption{Statistics derived from an application profile}
\vspace*{-4mm}
\centering
{\small
 \begin{tabular}{l  l  l} 
 \hline
 Notation & Description & Example\\
 \hline
 $N$ & \begin{tabular}{@{}l@{}}Containers per Node\end{tabular} & 1\\
 $M_{h}$ & \begin{tabular}{@{}l@{}}\poolheap\ size\end{tabular} & 4404MB\\
 \hline
 $CPU_{avg}$ & \begin{tabular}{@{}l@{}}Average CPU usage\end{tabular} & 35\%\\
 $Disk_{avg}$ & \begin{tabular}{@{}l@{}}Average disk usage\end{tabular} & 2\%\\
 \hline
 $M_{i}$ & \begin{tabular}{@{}l@{}}\poolcode\ 90\%ile value\end{tabular} & 115MB\\
 $M_{c}$ & \begin{tabular}{@{}l@{}}\poolcache\ 90\%ile value\end{tabular} & 2300MB\\
 $M_{s}$ & \begin{tabular}{@{}l@{}}\poolshuffle\ 90\%ile value\end{tabular} & 0MB\\
 $M_{u}$ & \begin{tabular}{@{}l@{}}\poolunmgd\ 90\%ile value\end{tabular} & 770MB\\
 \hline
 $P$ & \begin{tabular}{@{}l@{}}Task Concurrency\end{tabular} & 2\\
 $H$ & \begin{tabular}{@{}l@{}}Cache Hit Ratio (the fraction of cached\\ data partitions actually read from cache)\end{tabular} & 0.3\\
 $S$ & \begin{tabular}{@{}l@{}}Data Spillage Fraction (the fraction of\\ shuffle data spilled to disk)\end{tabular} & 0\\
 \hline
\end{tabular}\\
}
\vspace*{-6mm}
\label{tab:stats}
\end{table}

We use Thoth~\cite{thoth-action} framework to obtain a profile of the application. The profile includes the following:
\squishlist
 \item A timeline of the memory usage of JVM pools generated by the  JVM GC profiler~\cite{jmx} for every container
 \item A timeline of resource usage by every container generated by IBM's Performance Analysis Tool (PAT)~\cite{pat}
 \item A timeline of memory usage by application memory pools for cache and shuffle generated by custom instrumentation
 \item Application event log profile providing a timeline of tasks
\squishend

Table~\ref{tab:stats} lists the statistics derived from an application profile. The first two entries correspond to the container  configuration used. Values of these parameters are used by the Initializer module in estimating cache and shuffle memory requirements. Next, we obtain average CPU utilization and average disk utilization values from resource usage profiles. The requirement for \poolcode\ ($M_{i}$) is obtained by looking up heap usage value at the instance of the first task submission to the container. This value corresponds to the memory required for application code objects and is expected to be constant through the execution. The values obtained from multiple containers in an application profile could have a little variance, so we use a 90th percentile value for stability against outliers. The memory used by \poolcache\ ($M_{c}$) is computed by looking up the maximum cache usage value from the profile. The cache usage value may not necessarily correspond to the actual \poolcache\ requirement because the application could possibly have rejected some partitions from cache. We record Cache Hit Ratio ($H$) from application logs in order to evaluate the actual requirement. 


While both $M_{i}$ and $M_{c}$ are considered long term memory requirements of a container, the memory used for task execution ($M_{s} + M_{u}$), corresponds to short-term memory requirements. We assume that each running task equally contributes to the total task memory consumed in order to estimate \poolshuffle\ value ($M_{s}$). Like in the case of $M_{c}$, $M_{s}$ does not necessarily correspond to the actual \poolshuffle\ requirement of the application since the shuffle data could possibly have been spilled because of capacity constraints. Data spillage fraction ($S$) allows us to estimate the actual memory requirement. 
The \poolunmgd\ usage value is the hardest to obtain among the statistics presented in Table~\ref{tab:stats} since the application does not track this memory pool. We use JVM instrumentation to get a good estimate as detailed next.

As described in Section~\ref{sec:mem-overview}, JVM uses two garbage collection processes, namely, {\em young GC} and {\em full GC}, to collect any unreferenced objects from \poolheap. The {\em full GC} event cleans up garbage both from young generation and old generation pools. Monitoring heap usage right after a {\em full GC}, therefore, gives us a more accurate picture of task memory requirements. Referring to the pool organization shown in Figure~\ref{fig:applevel}, subtracting \poolcode\ memory and the instantaneous \poolcache\ value from the instantaneous \poolheap\ value gives us the memory used by tasks running at that instant. As stated previously, we assume that each running task contributes equally, and estimate the value of per task memory accordingly. Out of the two components in task memory, the instantaneous value of \poolshuffle\ is available from the instrumentation. The remaining component gives us the instantaneous \poolunmgd\ value. The 90th percentile over \poolunmgd\ values thus obtained at each {\em full GC} event gives us the final estimate of $M_{u}$.

\noindent \textit{ {\bf \em Example.} Statistics for the \pagerank\ application studied in Section~\ref{sec:reliability} are listed in the third column of Table~\ref{tab:stats}. It can be noticed that the application has a high \poolcache\ requirement indicated by a high $M_{c}$ and a low $H$. Further, a high $M_{u}$ indicates a high task memory footprint which makes the application susceptible to out-of-memory errors.}

\noindent {\bf Importance of full GC events}: 
In case the provided application profile contains no full GC events (significant of an application with very low memory footprint), estimating $M_{u}$ accurately becomes hard. One solution is to base the calculations on maximum \poolold\ pool occupancy. This approach, though, leads to an over-estimation of task memory requirements and in effect, {\em sub-optimal}, albeit {\em reliable} recommendations provided by the \relm\ tuner. We empirically study the sensitivity of recommendations to the provided profile in Section~\ref{sec:relm-eval}. 
The empirical analysis shows that the estimates made in absence of {\em full GC} events are off by up to two orders of magnitude. Based on this evidence, we discard using \poolold\ pool occupancy to estimate $M_{u}$. Instead, we recommend simple changes to the application configuration used for profiling. The changes are based on three practical heuristics for increasing GC pressure: (a) Decrease Heap Size, (b) Increase Task Concurrency, and (c) Increase NewRatio. The new profile generated using the heuristics is expected to contain {\em full GC} events, making it more suitable to the \relm\ tuner.

\mkcomment{Do we need to better handle the scenarios without full GC? It should not come across as a major weak point.}

\vspace*{-3mm}
\subsection{Initializer}
\label{sec:evaluator}
\vspace*{-1mm}

We use the statistics presented in Table~\ref{tab:stats} to configure each memory pool for a given container configuration identified by the Containers per Node $n$ and the Heap Size of each $m_{h}$. Notation of small letters is used to differentiate the test configuration from the profiled configuration used in statistics generation. A safety factor $\delta$ denotes a fraction of memory to be kept unassigned. It acts as a safeguard against {\em out-of-memory} errors. The Initializer uses  analytical models to configure each of the \poolcache, \poolshuffle, and \poolunmgd\ independently. Memory pressures and potential GC bottlenecks in the resulting configurations are handled by the {\em Arbitrator} module later. 

\noindent{\em Cache storage. }
\poolcache\ requirement is determined by scaling the maximum cache storage observed in the application profile by the cache hit ratio number. 
\begin{equation}
\vspace*{-2mm}
m_{c} = m_{h} * \text{min} \Big( \frac{M_{c}}{H * M_{h}}, 1-\delta \Big) \label{eq:cache}
\end{equation}

\noindent{\em Shuffle memory. }
We estimate \poolshuffle\ by scaling the maximum shuffle memory observed in the application profile by the data spillage fraction. It is assumed that each concurrently running task is an equal contributor to the spillage. 
\begin{equation}
\vspace*{-2mm}
m_{s} = \text{min} \Big( \frac{M_{s}}{1 - S / P}, (1-\delta) * m_{h} \Big) \label{eq:shuffle}
\end{equation}

\noindent{\em GC settings. }
The \poolold\ pool of JVM needs to be sized at least as big as the long term requirements, viz. $M_{i}$ and $m_{c}$, in order to lower the GC overheads (Section~\ref{sec:gc}). The GC parameter $NewRatio (NR)$ is set accordingly. \pooleden\ size is calculated by subtracting two survivor spaces specified by $SurvivorRatio (SR)$ from Young pool size. 
\begin{equation}
\vspace*{-2mm}
\begin{split}
NR &= \text{ceil} \Big( \frac{M_{i} + m_{c}}{m_{h} - M_{i} - m_{c}}\Big) \\
m_{o} = m_{h} * \frac{NR}{NR + 1},\  &
m_{e} = m_{h} * \frac{1}{NR + 1} * \frac{SR - 2}{SR}
\label{eq:gc}
\end{split}
\end{equation}

\noindent{\em Task concurrency. }
Number of tasks that can run concurrently in a container is estimated based on the following stats obtained from the application profile: (a) average CPU per task, (b) average disk usage per task, and (c) maximum per-task memory requirements. The models assume a linear relation to obtain a conservative estimate. 
\begin{equation}
\vspace*{-2mm}
\begin{split}
p^{CPU} = \frac{1}{n} \frac{(1-\delta) * 100} {CPU_{avg} / P},\ &
p^{disk} = \frac{1}{n} \frac{(1-\delta) * 100} {Disk_{avg} / P}\\
p^{memory} = \frac{(1 - \delta) * m_{h}} {M_{u}},\ &
p  = \text{min}(p^{CPU}, p^{disk}, p^{memory}) \label{eq:task}
\end{split}
\end{equation} 

\mkcomment{Need to convince why such simplistic models work.}

\noindent\textit{{\bf \em Example.} The \pagerank\ application studied in Section~\ref{sec:reliability} when evaluated on the container configuration of $n=1$ and $m_{heap}=4404MB$, with safety factor $\delta=0.1$, results in the following:}
\begin{equation}
\vspace*{-3mm}
m_{c}  = 3798 MB, 
m_{s} = 0 MB, 
p = 5, 
NR = 9
\label{eq:pagerank-init}
\end{equation}

\vspace*{-3mm}
\subsection{Arbitrator}
\label{sec:reliable}
\vspace*{-1mm}

\begin{figure*}
\centering
\begin{tikzpicture}
\begin{axis}[
title={(1)},
title style={yshift=-1.5ex},
ylabel={$p:5, m_{c}:3.7GB, NR:9$},
ylabel near ticks,
ylabel shift={-8pt},
height=4.8cm,
width=2.5cm,
legend columns=-1,
legend to name=named,
smooth, thick,
enlarge x limits={abs=0.25cm},
xtick={}, xticklabels={},
ytick={4404}, yticklabels={},
hide x axis,
axis y line*=left,
ymax=8000,
ybar stacked,
]
\addplot+[ybar, postaction={pattern=north east lines}] coordinates
{(1, 3963.6) (2, 0)};
\addplot+[ybar] coordinates
{(1, 0) (2, 122.0703125)};
\addplot+[ybar, postaction={pattern=crosshatch dots}] coordinates
{(1, 0) (2, 3798.45)};
\addplot+[ybar, postaction={pattern=north west lines}] coordinates
{(1, 0) (2, 3680.229187)};
\addplot+[red, scatter, only marks, mark=star, mark size=3] coordinates {(1, 440.6)};
\end{axis}
\end{tikzpicture}
\quad
\begin{tikzpicture}
\begin{axis}[
title={(2)},
title style={yshift=-1.5ex},
ylabel={$p:4, m_{c}:3.7GB, NR:9$},
ylabel near ticks,
ylabel shift={-8pt},
height=4.8cm,
width=2.5cm,
legend columns=-1,
legend to name=named,
smooth, thick,
enlarge x limits={abs=0.25cm},
xtick={}, xticklabels={},
ytick={4404}, yticklabels={},
hide x axis,
axis y line*=left,
ymax=8000,
ybar stacked,
]
\addplot+[ybar, postaction={pattern=north east lines}] coordinates
{(1, 3963.6) (2, 0)};
\addplot+[ybar] coordinates
{(1, 0) (2, 122.0703125)};
\addplot+[ybar, postaction={pattern=crosshatch dots}] coordinates
{(1, 0) (2, 3798.45)};
\addplot+[ybar, postaction={pattern=north west lines}] coordinates
{(1, 0) (2, 2944.18335)};
\addplot+[red, scatter, only marks, mark=star, mark size=3] coordinates {(1, 440.6)};
\end{axis}
\end{tikzpicture}
\quad
\begin{tikzpicture}
\begin{axis}[
title={(3)},
title style={yshift=-1.5ex},
ylabel={$p:4, m_{c}:3GB, NR:3$},
ylabel near ticks,
ylabel shift={-8pt},
height=4.8cm,
width=2.5cm,
legend columns=-1,
legend to name=named,
smooth, thick,
enlarge x limits={abs=0.25cm},
xtick={}, xticklabels={},
ytick={4404}, yticklabels={},
hide x axis,
axis y line*=left,
ymax=8000,
ybar stacked,
]
\addplot+[ybar, postaction={pattern=north east lines}] coordinates
{(1, 3303) (2, 0)};
\addplot+[ybar] coordinates
{(1, 0) (2, 122.0703125)};
\addplot+[ybar, postaction={pattern=crosshatch dots}] coordinates
{(1, 0) (2, 3062.404163)};
\addplot+[ybar, postaction={pattern=north west lines}] coordinates
{(1, 0) (2, 2944.18335)};
\addplot+[red, scatter, only marks, mark=star, mark size=3] coordinates {(1, 1101)};
\end{axis}
\end{tikzpicture}
\quad
\begin{tikzpicture}
\begin{axis}[
title={(4)},
title style={yshift=-1.5ex},
ylabel={$p:4, m_{c}:3GB, NR:9$},
ylabel near ticks,
ylabel shift={-8pt},
height=4.8cm,
width=2.5cm,
legend columns=-1,
legend to name=named,
smooth, thick,
enlarge x limits={abs=0.25cm},
xtick={}, xticklabels={},
ytick={4404}, yticklabels={},
hide x axis,
axis y line*=left,
ymax=8000,
ybar stacked,
]
\addplot+[ybar, postaction={pattern=north east lines}] coordinates
{(1, 3963.6) (2, 0)};
\addplot+[ybar] coordinates
{(1, 0) (2, 122.0703125)};
\addplot+[ybar, postaction={pattern=crosshatch dots}] coordinates
{(1, 0) (2, 3062.404163)};
\addplot+[ybar, postaction={pattern=north west lines}] coordinates
{(1, 0) (2, 2944.18335)};
\addplot+[red, scatter, only marks, mark=star, mark size=3] coordinates {(1, 440.6)};
\end{axis}
\end{tikzpicture}
\quad
\begin{tikzpicture}
\begin{axis}[
title={(5)},
title style={yshift=-1.5ex},
ylabel={$p:3, m_{c}:3GB, NR:9$},
ylabel near ticks,
ylabel shift={-8pt},
height=4.8cm,
width=2.5cm,
legend columns=-1,
legend to name=named,
smooth, thick,
enlarge x limits={abs=0.25cm},
xtick={}, xticklabels={},
ytick={4404}, yticklabels={},
hide x axis,
axis y line*=left,
ymax=8000,
ybar stacked,
]
\addplot+[ybar, postaction={pattern=north east lines}] coordinates
{(1, 3963.6) (2, 0)};
\addplot+[ybar] coordinates
{(1, 0) (2, 122.0703125)};
\addplot+[ybar, postaction={pattern=crosshatch dots}] coordinates
{(1, 0) (2, 3062.404163)};
\addplot+[ybar, postaction={pattern=north west lines}] coordinates
{(1, 0) (2, 2208.137512)};
\addplot+[red, scatter, only marks, mark=star, mark size=3] coordinates {(1, 440.6)};
\end{axis}
\end{tikzpicture}
\quad
\begin{tikzpicture}
\begin{axis}[
title={(6)},
title style={yshift=-1.5ex},
ylabel={$p:3, m_{c}:2.3GB, NR:2$},
ylabel near ticks,
ylabel shift={-8pt},
height=4.8cm,
width=2.5cm,
legend columns=-1,
legend to name=named,
smooth, thick,
enlarge x limits={abs=0.25cm},
xtick={}, xticklabels={},
ytick={4404}, yticklabels={},
hide x axis,
axis y line*=left,
ymax=8000,
ybar stacked,
]
\addplot+[ybar, postaction={pattern=north east lines}] coordinates
{(1, 2936) (2, 0)};
\addplot+[ybar] coordinates
{(1, 0) (2, 122.0703125)};
\addplot+[ybar, postaction={pattern=crosshatch dots}] coordinates
{(1, 0) (2, 2326.16759)};
\addplot+[ybar, postaction={pattern=north west lines}] coordinates
{(1, 0) (2, 2208.137512)};
\addplot+[red, scatter, only marks, mark=star, mark size=3] coordinates {(1, 1468)};
\end{axis}
\end{tikzpicture}
\quad
\begin{tikzpicture}
\begin{axis}[
title={(7)},
title style={yshift=-1.5ex},
ylabel={$p:3, m_{c}:2.3GB, NR:6$},
ylabel near ticks,
ylabel shift={-8pt},
height=4.8cm,
width=2.5cm,
legend columns=-1,
legend to name=named,
smooth, thick,
enlarge x limits={abs=0.25cm},
xtick={}, xticklabels={},
ytick={4404}, yticklabels={},
hide x axis,
axis y line*=left,
ymax=8000,
ybar stacked,
]
\addplot+[ybar, postaction={pattern=north east lines}] coordinates
{(1, 3774.857143) (2, 0)};
\addplot+[ybar] coordinates
{(1, 0) (2, 122.0703125)};
\addplot+[ybar, postaction={pattern=crosshatch dots}] coordinates
{(1, 0) (2, 2326.16759)};
\addplot+[ybar, postaction={pattern=north west lines}] coordinates
{(1, 0) (2, 2208.137512)};
\addplot+[red, scatter, only marks, mark=star, mark size=3] coordinates {(1, 629.1428571)};
\end{axis}
\end{tikzpicture}
\quad
\begin{tikzpicture}
\begin{axis}[
title={(8)},
title style={yshift=-1.5ex},
ylabel={$p:2, m_{c}:2.3GB, NR:6$},
ylabel near ticks,
ylabel shift={-8pt},
height=4.8cm,
width=2.5cm,
legend columns=-1,
legend to name=named,
smooth, thick,
enlarge x limits={abs=0.25cm},
xtick={}, xticklabels={},
ytick={4404}, yticklabels={},
hide x axis,
axis y line*=left,
ymax=8000,
ybar stacked,
]
\addplot+[ybar, postaction={pattern=north east lines}] coordinates
{(1, 3774.857143) (2, 0)};
\addplot+[ybar] coordinates
{(1, 0) (2, 122.0703125)};
\addplot+[ybar, postaction={pattern=crosshatch dots}] coordinates
{(1, 0) (2, 2326.16759)};
\addplot+[ybar, postaction={pattern=north west lines}] coordinates
{(1, 0) (2, 1472.091675)};
\addplot+[red, scatter, only marks, mark=star, mark size=3] coordinates {(1, 629.1428571)};
\end{axis}
\end{tikzpicture}
\quad
\begin{tikzpicture}
\begin{axis}[
title={(9)},
title style={yshift=-1.5ex},
ylabel={$p:2, m_{c}:1.5GB, NR:1$},
ylabel near ticks,
ylabel shift={-8pt},
height=4.8cm,
width=2.5cm,
legend columns=-1,
legend to name=named,
smooth, thick,
enlarge x limits={abs=0.25cm},
xtick={}, xticklabels={},
ytick={4404}, yticklabels={},
hide x axis,
axis y line*=left,
ymax=8000,
ybar stacked,
]
\addplot+[ybar, postaction={pattern=north east lines}] coordinates
{(1, 2202) (2, 0)};
\addplot+[ybar] coordinates
{(1, 0) (2, 122.0703125)};
\addplot+[ybar, postaction={pattern=crosshatch dots}] coordinates
{(1, 0) (2, 1590.121753)};
\addplot+[ybar, postaction={pattern=north west lines}] coordinates
{(1, 0) (2, 1472.091675)};
\addplot+[red, scatter, only marks, mark=star, mark size=3] coordinates {(1, 2202)};
\end{axis}
\end{tikzpicture}
\quad
\begin{tikzpicture}
\begin{axis}[
title={(10)},
title style={yshift=-1.5ex},
ylabel={$p:2, m_{c}:1.5GB, NR:3$},
ylabel near ticks,
ylabel shift={-8pt},
height=4.8cm,
width=2.5cm,
legend columns=-1,
legend entries={$m_{o}\ $, $M_{i}\ $, $m_{c}\ $, $p * M_{u}\ $, $m_{h}\ $},
legend to name=named,
smooth, thick,
enlarge x limits={abs=0.25cm},
xtick={}, xticklabels={},
ytick={4404}, yticklabels={},
hide x axis,
axis y line*=left,
ymax=8000,
ybar stacked,
]
\addplot+[ybar, postaction={pattern=north east lines}] coordinates
{(1, 3303) (2, 0)};
\addplot+[ybar] coordinates
{(1, 0) (2, 122.0703125)};
\addplot+[ybar, postaction={pattern=crosshatch dots}] coordinates
{(1, 0) (2, 1590.121753)};
\addplot+[ybar, postaction={pattern=north west lines}] coordinates
{(1, 0) (2, 1472.091675)};
\addplot+[red, scatter, only marks, mark=star, mark size=3] coordinates {(1, 1101)};
\end{axis}
\end{tikzpicture}
\\
\ref{named}
\vspace*{-4mm}
\caption{Working example showing steps of \relm's {\em Arbitrator} algorithm on \pagerank\ application}
\vspace*{-4mm}
 \label{fig:example}
\end{figure*}
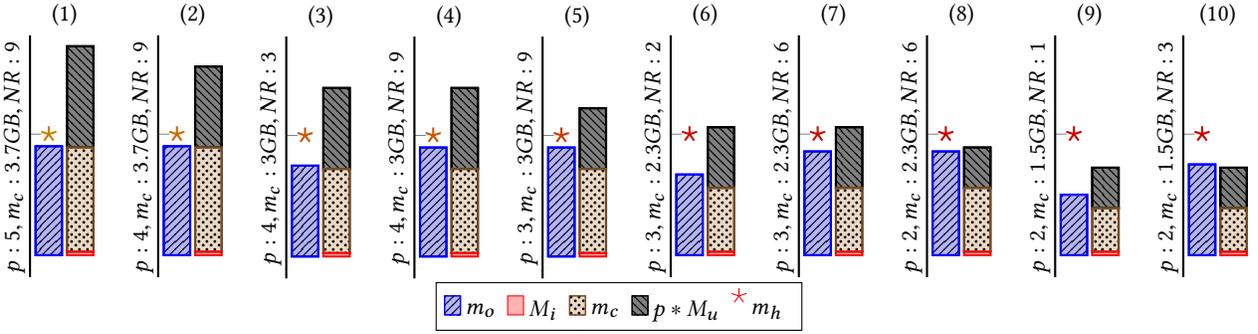

\noindent Building on the observations made in the empirical analysis, we present a general algorithm to tune a given configuration for reliability and low GC overheads. Algorithm~\ref{alg:reliable} presents the pseudo-code.

\vspace*{-3mm}
\begin{algorithm}
\caption{{Arbitrator}}\label{alg:reliable}
\hspace*{\algorithmicindent} \textbf{Input:} Configuration $\mb{c} = (M_{i}, M_{u}, p, m_{c}, m_{s})$, Safety factor $\delta$\\
 \begin{algorithmic}[1]
 \If {$(M_{i} + M_{u}) > (1-\delta) * m_{h}$}
  \State Return flagging insufficient memory
 \EndIf
 \While {$(M_{i} + p * M_{u} + m_{c}) > m_{o}$}
   \State one of the following three in a round-robin manner:
   \State {\bf I.} Decrease $p$ by 1 if $p > 1$
   \State {\bf II.} Reduce $m_{c}$ by $M_{u}$ ensuring that $m_{c} > 0$.\\\hspace*{8mm} Change GC pools using Equation~\ref{eq:gc}.
   \State {\bf III.} Increase $m_{o}$ by $M_{u}$ ensuring that $m_{o} < (1-\delta) * m_{h}$
  \EndWhile
 \State Set shuffle memory $m_{s} = \text{min} (m_{s}, 0.5 * m_{e} / p)$
 \State Set output $\mb{C} = (M_{i}, M_{u}, p, m_{c}, m_{s})$
 \State Set utility score $\rmv_{\mb{C}} = \frac{M_{i} + m_{c} + p * ( M_{u} + m_{s} )}{m_{h}}$
 \State Return ($\mb{C}$, $\rmv_{\mb{C}}$).
  \end{algorithmic}
\end{algorithm}
\vspace*{-4mm}

Line 1 checks if the configuration satisfies the bare minimum requirement of a container running at least one task at any given time. Lines 4-10 represent the main loop where actions to change configuration are carried out if the combined memory consumed by \poolcode, \poolcache, and \poolunmgd\ exceeds \poolold. Please recall that the task memory values are obtained by profiling {\em full GC} events and correspond to the task objects tenured to \poolold. If the combined memory exceeds $m_{o}$, we perform one of the three actions given in Lines 6, 7, and 9 in a round-robin manner:
\squishlist
 \item Decrease Task Concurrency by 1. This reduces the memory footprint by $M_{u}$.
 \item Decrease Cache Capacity by $M_{u}$. We also adjust GC pools so that \poolold\ pool is just larger than the value $M_{i} + m_{c}$. The idea is to probe if an optimal GC setting for the given \poolcache\ value can ensure safety as well.
 \item Increase old generation pool size by $M_{u}$. This optimization trades-off performance to ensure safety against out-of-memory errors (Recall {\em Observation 6} from Section~\ref{sec:gc}). 
\squishend

At the end of the loop, settings for Task Concurrency, Cache Capacity, and NewRatio are locked in. Based on the available \pooleden, \poolshuffle\ is tuned in Line 11 which avoids the high GC overheads explained in Figure~\ref{fig:sortgc}. Finally, Line 13 computes a utility score \rmv\ which corresponds to the fraction of \poolheap\ allocated to the internal memory pools.

\noindent \textit{{\bf \em Example.} Continuing with the \pagerank\ example for which the configuration produced by the initializer is given in Eq.~\ref{eq:pagerank-init}. Figure~\ref{fig:example} details the changes in memory pools starting with the initial configuration shown in (1). It takes 9 iterations of the main loop to reach a reliable configuration which sets Task Concurrency = 2, Cache Capacity = 1.5GB, and NR = 3. Compared to the profiled application run, in which containers fail with out-of-memory errors, this configuration lowers the cache capacity by 700MB per container 
thereby improving reliability of the application. This, however, is not the only reliable configuration \relm\ finds: A better performing configuration is obtained when the process is repeated on a configuration of 2 Containers per Node. Section~\ref{sec:comparisons} presents this result.
}

\noindent{\bf Analysis: } As stated in \relm\ goals, {\em safety} is the primary objective. The Arbitrator meets this objective by ensuring that the combined allocation of internal memory pools remains within \poolheap. The next two performance objectives, a {\em high task concurrency} and a {\em high cache hit ratio}, are achieved by a two-phase process. {\em Initializer} first optimizes the \poolunmgd\ and \poolcache\ pools corresponding to the two requirements independently against the entire heap size. {\em Arbitrator} then takes small chunks out of the two pools in a round-robin manner until it can meet the {\em safety} condition. This process results in a {\em proportionally fair}~\cite{proportional-fair} allocation for the two memory pools. The arbitration is invoked for each enumerated container configuration which is a small number because of the physical constraints in resource allocation. Within an invocation, the number of iterations of the main loop is a linear function of the maximum degree of parallelism (number of cores) in the worst case. So overall, the algorithm needs only a handful steps to recommend a configuration that best meets the goals (1), (2a). (2b). and (3).

\vspace*{-3mm}
\section{Black-box tuners}
\label{sec:ai}

AI-driven black-box formulation is a popular choice for auto-tuning because of its applicability to a wide variety of problem setups. The basic idea is to incrementally probe samples from the space of configuration options to learn their impact on performance. 
Approaches differ in terms of how they explore the configuration space. As an example, Elastisizer~\cite{elastisizer}, a tool to auto-tune cluster sizes for cloud platforms, uses {\em Recursive Random Search}~\cite{rrs} which samples the search space randomly to find promising regions to recursively probe into. We adopt two popular techniques to our problem: (1) A sequential model-based optimization called {\em Bayesian Optimization}, and (2) A model-free deep reinforcement learning algorithm called {\em Deep Distributed Policy Gradient}.

\vspace*{-3mm}
\subsection{Bayesian Optimization}
\label{sec:bo}
\vspace*{-1mm}

The state-of-the-art black-box optimization approaches~\cite{ituned, ottertune, cherrypick} use {\em Bayesian Optimization}~\cite{bayesian-book} which is a powerful learning technique with power equivalent to that of deep networks. It allows us to approximate complex response surfaces through adaptive sampling of the search space in a manner which balances exploration (i.e.,  probing new regions) and exploitation (i.e., favoring the promising regions). At the core of the \bayesian\ is a surrogate model used to approximate the response surface. {\em Gaussian Process}~\cite{gaussian} is an attractive choice for the surrogate model because of its salient features such as confidence bound on predictions, support for noisy observations, and an ability to use gradient-based methods~\cite{ieee16}. 

Alternate surrogate models such as Random Forest and Boosted Regression Trees have been shown to be better at modeling the non-linear interactions~\cite{arrow}. However, they lack theoretical guarantees on the confidence bounds that Gaussian Process offers. Also we did not find much qualitative difference among the models when evaluated in our setup and, therefore, do not include them in the discussion here.

We model our problem using the Gaussian Process next. 
We are given a data analytics application $A$ and $d$ parameters $x_1, x_2, \ldots, x_d$ to tune. The parameters correspond to the options used to control usage of various memory pools as listed in Table~\ref{tab:options}. 
The performance metric, denoted by $y$, corresponds to the wall-clock duration of the application $A$ on a setting $(x_1, x_2, \ldots, x_d) \in \mc{X}$.
Tuning is carried out by adaptively collecting samples $\langle \mb{x},y \rangle = \langle x_1=v_1, x_2=v_2, \ldots, x_d=v_d, y=p \rangle$. 
The prior belief in Gaussian Process is modeled as $f(\mathbf{x}) \sim GP(\mu_0, k)$, 
where $\mu_0:\mc{X} \rightarrow \mathbb{R}$ denotes the prior mean function 
and $k:\mc{X} \times \mc{X} \rightarrow \mathbb{R}$ denotes the covariance function. 
Given $n$ sampled points $\mb{x}_{1:n}$ and noisy observations $y_{1:n}$ ($\sigma^2$ denoting a constant observation noise), 
the unknown function values $\mathbf{f} := f_{1:n}$ are assumed to be jointly Gaussian, i.e.~$\mb{f}|\mb{x} \sim \mc{N}(\mb{m}, \mb{K})$, 
and the observations $\mb{y} := y_{1:n}$ are normally distributed given $\mathbf{f}$, i.e.~$\mb{y}|\mb{f},\sigma^2 \sim \mc{N}(\mb{f}, \sigma^2\mb{I})$. The posterior mean and variance are then given by the following:

\vspace*{-5mm}
\begin{equation}
\begin{split}
\mu_n(\mb{x}) &= \mu_0(\mb{x}) + \mb{k}(\mb{x})^\top (\mb{K} + \sigma^2\mb{I})^{-1}(\mb{y}-\mb{m})\\
\sigma^2_n(\mb{x}) &= k(\mb{x}, \mb{x}) - \mb{k}(\mb{x})^\top(\mb{K}+\sigma^2\mb{I})^{-1}\mb{k}(\mb{x})
\vspace*{-6mm}
\end{split}
\end{equation}

\noindent where $\mb{k}(\mb{x})$ is a vector of covariance between $\mb{x}$ and $\mb{x}_{1:n}$.

An acquisition function provided by \bayesian\ suggests the next probe based on the posterior distribution. We use one of the most popular acquisition functions, Expected Improvement (EI), given below:

\vspace*{-4mm}
\begin{equation}
EI(\mb{x}; \mb{x}_{1:n}, y_{1:n}) = (\tau - \mu_n(\mb{x})) \Phi(Z) + \sigma_n(\mb{x}) \phi(Z)
\label{eq:ei}
\vspace*{-1mm}
\end{equation}

\noindent Here, $\tau$ denotes the current best observation, $Z = (\tau - \mu_n(\mb{x})) / \sigma_n(\mb{x})$, and $\Phi$ and $\phi$ are the standard normal cumulative distribution and density functions respectively. The next sample will be either picked from a region where uncertainty is high, captured by $\sigma_n(\mb{x})$, or from a region close to the current best, captured by $(\tau - \mu_n(\mb{x}))$, thus balancing the {\em exploration} and the {\em exploitation}. A combination of random sampling and standard gradient-based search is carried out to find the highest expected improvement.

The number of samples needed for \gaussian, or for any regression technique in general, can be very high if the number of independent configuration knobs is high. Therefore, tuning is often preceded by a feature selection phase to identify a subset of important features that significantly affect the performance. For example, OtterTune~\cite{ottertune} uses {\em Lasso}~\cite{lasso} technique to cut down the number of tuning knobs for DBMSs from hundreds down to a handful. The tuning parameters in our setup, however, all have a significant performance impact as we show in Section~\ref{sec:interactions}, making the feature selection redundant.

In our implementation of \gaussian, we start with a small number of samples taken using Latin Hypercube Sampling (LHS)~\cite{lhs} over the domain space $\Pi^d_{i=1} dom(x_i)$. LHS is an efficient technique to generate near-random samples from a multidimensional space while providing a good coverage. These samples initialize the Gaussian process. We continue taking more samples adaptively as suggested by the \gaussian\ until the expected improvement falls below a 10\% threshold and at least 6 new configurations have been observed; this stopping condition, borrowed from CherryPick~\cite{cherrypick}, is developed to give sufficient chance for the black-box optimization policy to generate a decent recommendation. 


\vspace*{-3mm}
\subsection{Guided Bayesian Optimization}
\label{sec:guided}
\vspace*{-1mm}

Bayesian Optimization, being a black-box policy, often requires a number of sample runs to develop sufficient confidence in its predictions. Recent work has shown that using execution profiles of applications along with knowledge of system internals can help speed up the tuning process significantly. Dalibard et.~al.~\cite{boat} propose Structured Bayesian Optimization (SBO) which lets system developers develop bespoke probabilistic models by including simple parametric models inferred from low-level performance metrics observed during a tuning run. The combination of non-parametric bayesian optimizer and the evolving parametric models helps achieve a faster convergence compared to a vanilla Bayesian Optimizer. Arrow~\cite{arrow}, targeted at finding best VM configurations, augments a bayesian optimizer driven by VM characteristics with low-level performance metrics for the same purpose. Following in with the same philosophy, we design Guided Bayesian Optimization (\guided) and deploy it in tuning memory-based analytics applications.

\begin{figure}[]
\centering
 \includegraphics[width=0.8\columnwidth]{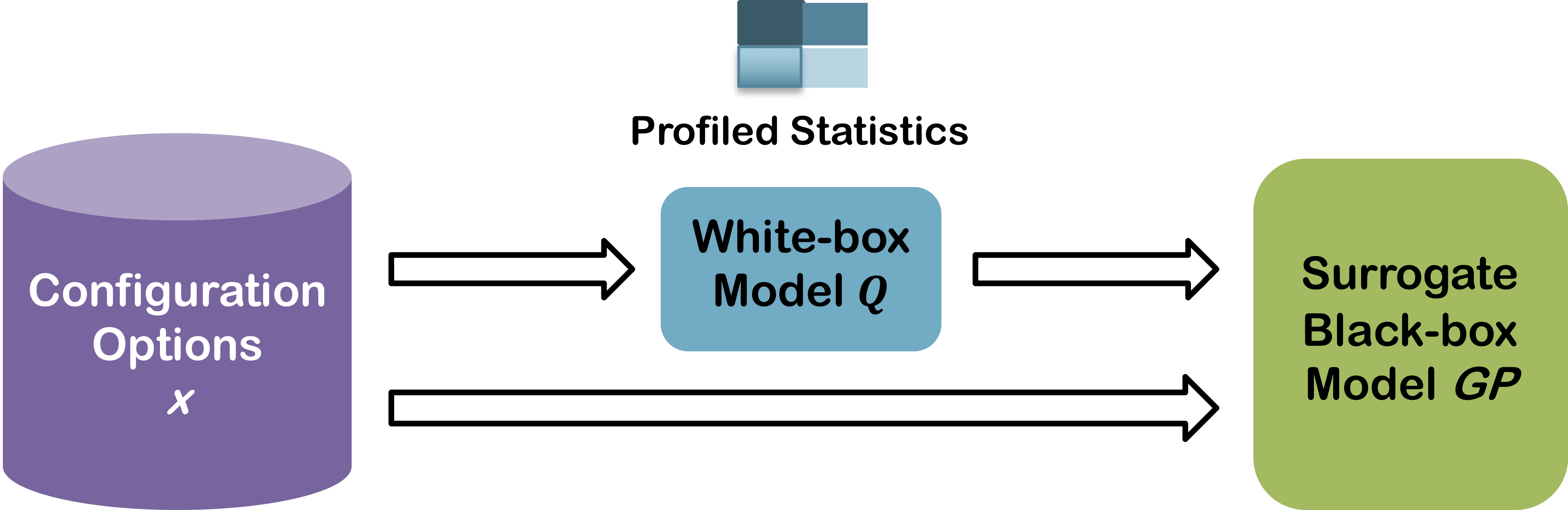}
\vspace*{-3mm}
\caption{Design of Guided Bayesian Optimization (\guided)}
\vspace*{-7mm}
\label{fig:gboflow}
\end{figure}

Figure~\ref{fig:gboflow} shows the concept of \guided. The most important building block of \guided\ is a white-box model which is given a configuration and a set of profiled statistics for the application under test. The model outputs a set of derived metrics which are used in addition to the original configuration options for the optimization. The additional metrics are derived using simple analytical models with the purpose of separating out the most suitable region of configuration space from the more expensive region. Compared to SBO~\cite{boat}, which requires a system expert to design a parametric model by observing the system performance while tuning, \guided\ simplifies the process with a white-box model that can be used right from the start of the optimization process on any type of workload. 

\noindent{\bf Guiding white-box model:}
The model used for guiding exploration ($Q$) is based on the empirical analysis carried out in Section~\ref{sec:interactions}. Inputs to the model include: (a) Configuration options under test ($\mb{x}$), and (b) Profiled statistics from a prior execution, not necessarily using the same configuration (Table~\ref{tab:stats}).
\begin{equation}
\vspace*{-1mm}
\begin{split}
q_1^\mb{x} &= \frac{M_i + \text{min}(m_c^\mb{x}, m_c) + p^\mb{x} * (M_u + \text{min}(m_s^\mb{x}, m_s)}{m_h^\mb{x}}) \\ 
q_2^\mb{x} &= \frac{M_i + m_c}{\text{min}(m_o^\mb{x}, m_c^\mb{x})}\ ,\ 
q_3^\mb{x} = \frac{p^\mb{x} * \text{min}(m_s^\mb{x}, m_s)}{0.5*m_e^\mb{x}}\\
\mb{q}^\mb{x} &= \{q_1^\mb{x}, q_2^\mb{x}, q_3^\mb{x}\} \label{eq:q}
\end{split}
\vspace*{-3mm}
\end{equation}

$Q$ generates three metrics as listed in Eq.~\ref{eq:q}. $q_1$ corresponds to the expected heap occupancy of a container. The numerator adds up the expected memory usage by every application level memory pool. The \poolcache\ and \poolshuffle\ requirements (denoted by $m_c$ and $m_s$) are modeled by Eq.~\ref{eq:cache} and Eq.~\ref{eq:shuffle} respectively. The intuition is to identify both the configurations under-utilizing memory (those with low scores) as well as the potentially {\em unsafe} ones (those with scores over 1). $q_2$ corresponds to the expected long term memory efficiency. Here, the numerator corresponds to the long term memory requirement while the denominator corresponds to the available long term memory storage considering the limits enforced by the configuration options. A high $q_2$ score could mean either high disk overheads on account of data not fitting in memory or high GC overheads on account of data not fitting in \poolold\ pool (Recall {\em Observation 5} from Section~\ref{sec:interactions}). $q_3$ corresponds to the efficiency of the shuffle memory usage. Based on {\em Observation 7}, a high $q_3$ score means high GC overheads because of the large-sized data spills.

The set of metrics derived by model $Q$ is designed to be the most practical means to identify safe, highly efficient, and low overhead configurations in accordance with the goals set out by \relm. This set could be expanded to add more indicators of the \relm\ goals. We plan to work on supporting a mechanism to add more metrics while ensuring that they form an independent set of features and are ranked as per their importance to the estimation.

 
\noindent{\bf Changes to surrogate model:}
The surrogate model used in Section~\ref{sec:ai} uses a Gaussian Process to fit in existing observations and can be represented as $GP(\mb{x}_{1:n}, y_{1:n})$. With the additional inputs coming from the white-box model $Q$, the model is now modified to $GP(\mb{x}_{1:n}, \mb{q}_{1:n}, y_{1:n})$. As described earlier, the Bayesian optimizer uses a small number of uniform random samples and a few invocations of quasi-Newton hill climbers (e.g.~L-BFGS~\cite{l-bfgs}) to explore the space of unseen configurations ($\mb{X}$). The next probe is identified using the Expected Improvement score as formalized next.
\vspace*{-2mm}
\begin{equation}
\mb{x}_{n+1} = \argmax_{\mb{x} \in \mb{X}}EI(\mb{x}, \mb{q}^\mb{x}; \mb{x}_{1:n}, \mb{q}_{1:n}, y_{1:n}) \label{eq:probe}
\vspace*{-1mm}
\end{equation}

\vspace*{-4mm}
\subsection{Reinforcement Learning}
\label{sec:rl}
\vspace*{-1mm}

Reinforcement Learning (RL) involves an agent that interacts with an environment $E$ in discrete timesteps. At each timestep $t$, it makes an observation, takes an action $a_t$, and receives a reward $r_t$. The action changes the state of the environment to $s_t$. 
We first map the terminology to our setup before describing the specific RL agent we use.

\begin{figure}[]
\centering
 \includegraphics[width=0.7\columnwidth]{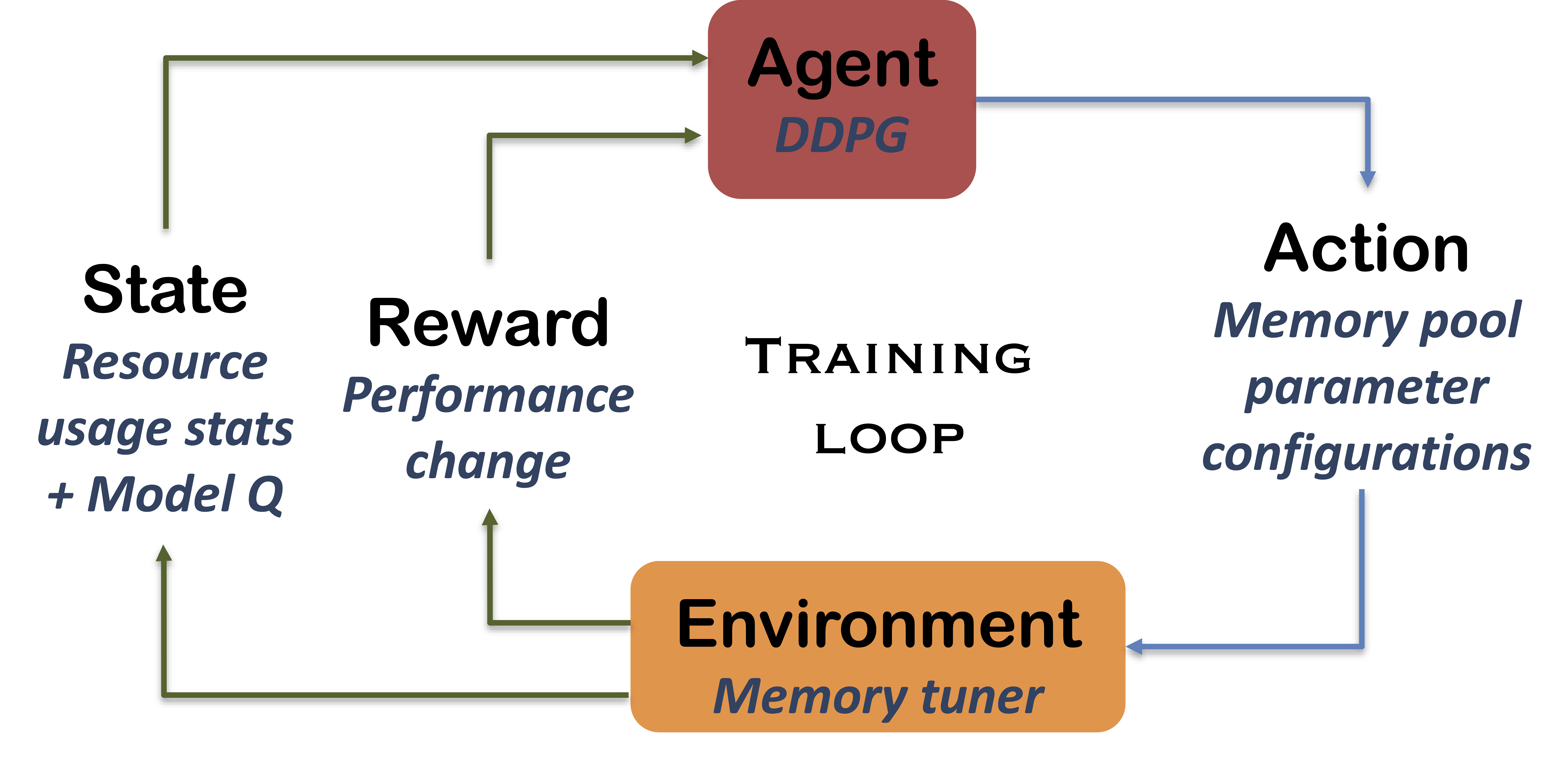}
\vspace*{-4mm}
\caption{Reinforcement learning adopted for auto-tuning memory-based analytics}
\vspace*{-5mm}
\label{fig:ddpg}
\end{figure}

Figure~\ref{fig:ddpg} shows the adoption of RL for the problem of tuning a given data analytics application. An {\em action} constitutes a change in configuration knobs (listed in Table~\ref{tab:options}). Similar to the approach used in CDBTune~\cite{cdbtune} for DBMS tuning, a {\em state} corresponds to a set of resource usage metrics. The statistics on CPU, IO, and memory usage listed in Table~\ref{tab:stats} constitute one half of the metrics. Following the philosophy of \guided, we add to this set the metrics derived from model $Q$ (Eq.~\ref{eq:q}) to get a visibility into utilization of the internal memory pools. The {\em reward} function is borrowed from CDBTune as well; it considers the performance change at not only the previous timestep but also considers the first timestep when the tuning request was made.

\noindent{\bf DDPG Overview:}
Deep Deterministic Policy Gradient~\cite{ddpg} is a policy-based model-free RL agent which combines Deep Q Neural Network with Actor-Critic models to work with continuous configuration parameter space. 
The \reinforce\ actor learns a policy function $a_t = \mu(s_t | \theta^\mu)$, where $\theta^\mu$ maps state $s_t$ to value of action $a_t$. A critic function $\mathcal{Q}(s_t, a_t | \theta^{\mathcal{Q}})$ evaluates the policy function estimated by the actor. Evaluation of the value considers not only the current reward but also discounted future rewards. \reinforce\ uses an {\em experience replay} memory to store the explored state-action pairs and uses a sample from the memory for learning its critic model. 

\reinforce, being a model-free algorithm, does not need to store all the combinations of states and actions it has explored. Exploration of action space is carried out by adding a noise sampled from a noise process $\mathcal{N}$ to the actor $\mu$. Details of \reinforce\ algorithm are not included here, but can be found in~\cite{ddpg}.

\vspace*{-2mm}
\section{Evaluation}
\label{sec:eval}

\vspace*{-1mm}
\subsection{Setup}
\label{sec:setup}
\vspace*{-1mm}

Our evaluation uses two Spark clusters listed in Table~\ref{tab:setup}. The applications we have picked for evaluation represent Map and Reduce computations, machine learning, distributed graph processing, and SQL processing use cases. 
The test suite including input data sources is provided in Table~\ref{tab:benchmarks}. The input data is stored in HDFS co-located with the compute cluster. We have deliberately changed partition sizes for some of the applications (namely, \sort\ and \svm) from the default HDFS block size of 128MB to create another dimension of variability in the test suite.


\noindent{\bf Configuration Space. }
The configuration options we tune correspond to the parameters controlling memory pools listed in Table~\ref{tab:options}.

The maximum heap available for allocation per node is 4404MB on cluster A and 16GB on cluster B.  We allow it to be distributed equally among 1, 2, 3, or 4 containers per node creating four possible configurations.
The number of concurrently running tasks on a node is limited by the number of physical cores. Therefore, the Task Concurrency value can range from 1 to the ratio of the physical cores to the number of containers. For example, if 2 containers are launched on a node with 8 physical cores, Task Concurrency on each container ranges from 1 to 4.

Cache Capacity and Shuffle Capacity values are set as a fraction (ranging from 0 to 1) of \poolheap. As Spark provides a unified memory pool~\cite{unified} combining both \poolcache\ and \poolshuffle, we set the capacity of the unified pool to the sum of Cache Capacity and Shuffle Capacity. 

When it comes to the GC parameters, the lowest possible value for NewRatio is 1. The maximum, while unbounded in theory, is limited to 9 in our setup. Higher values for NewRatio lead to too many invocations of GC because of the very low capacity of the young generation pool. Our heuristic of capping the value of NewRatio to 9 ensures that at least 10\% of Heap is available to the young generation pool. We keep the SurvivorRatio to its default value.

\noindent{\bf Default Policy:}\\
The default configuration by Amazon EMR's \maximizeresource\ policy~\cite{spark-emr} is listed in Table~\ref{tab:default}. This policy starts a single container on each node allocating it all the memory and the compute resources available on the node. Since the cluster A closely mimics the hardware configuration of Amazon EC2 node types {\em m4.large}, the default policy starts each container with a Heap Size of 4400MB and Task Concurrency set to 2. These settings do not vary across applications.

\noindent{\bf Exhaustive Search. }
Our exhaustive search policy grids the configuration space by discretizing the Domain of each parameter into 4 values 
We use only one of Cache Capacity and Shuffle Capacity depending on the dominant requirement of the application under test just to avoid collecting insignificant data. The minor memory pool capacity is set to 0.1. Despite the dimensionality reduction, \exhaustive\ is clearly an inefficient policy: The time taken to run all 192 configurations for an application on cluster A is at least 3 days. We performed the \exhaustive\ only in order to compare the quality of results produced by the other tuning policies.

\noindent{\bf Black-box Policy. }
As detailed in Section~\ref{sec:ai}, we use Bayesian Optimization as our candidate for black-box tuning. \bayesian\ is implemented using {\em scikit-learn} library in Python~\cite{scikit}. Like in the case of \exhaustive, only the dominant memory pool between \poolcache\ and \poolshuffle\ is used for optimization, with the minor pool capacity set to 0.1. Since the accuracy of \bayesian\ predictions depends on the number of samples explored, we bootstrap the model with 4 samples generated using Latin Hypercube Sampling~\cite{lhs} as listed in Table~\ref{tab:lhs}. The number 4 corresponds to the dimension of the configuration space we have used in the evaluation. LHS provides a good coverage of the configuration space setting up good priors for \gaussian\ search. 

The objective function is set to the application runtime. If a run is aborted due to errors, the objective value for the sample is set to twice the worst runtime obtained on the samples explored so far. This heuristic ensures that the failing region is ranked low during exploration. The same setup is mimicked for our optimized policy of Guided Bayesian Optimization (\guided).

Reinforcement learning (\reinforce) is another black-box policy we evaluate. DDPG algorithm described in Section~\ref{sec:rl} is implemented using PyTorch~\cite{pytorch} library with its neural network parameters borrowed from CDBTune~\cite{cdbtune}.

\noindent{\bf White-box Policy. }
\relm\ is our white-box model. It uses Thoth~\cite{thoth} framework to collects application profiles with minimal overheads. The Modules {\em Initializer} and {\em Arbitrator} are implemented in Java with $\approx 200$ lines of code; the source is available online~\cite{relm-source}. The safety fraction $\delta$ is set to 0.1 throughout the evaluation.

We first compare the quality of results of all the aforementioned policies in Section~\ref{sec:comparisons} before analyzing their overheads in Section~\ref{sec:overheads}. The \relm\ model and the \guided\ model are analyzed separately in Section~\ref{sec:relm-eval} and Section~\ref{sec:bo-eval} respectively. Finally, we briefly discuss general applicability of our models to the changes in data and/ or hardware in Section~\ref{sec:reuse}.

\vspace*{-2mm}
\subsection{Quality of Results}
\label{sec:comparisons}
\vspace*{-1mm}

\begin{table}
\centering
\caption{Samples generated by Latin Hypercube Sampling used in \gaussian\ initialization.}
\vspace*{-4mm}
\resizebox{0.7\columnwidth}{!}{%
 \begin{tabular}{|c c c c|}
 \hline
 \begin{tabular}{@{}c@{}}Containers\\ per Node\end{tabular} & \begin{tabular}{@{}c@{}}Task\\ Concurrency\end{tabular} & \begin{tabular}{@{}c@{}}Cache Capacity/\\ Shuffle Capacity\end{tabular} & NewRatio\\
 \hline
 1 & 4 & .6 & 7 \\
 2 & 1 & .4 & 3 \\
 3 & 2 & .2 & 5 \\
 4 & 2 & .8 & 1 \\
 \hline
 \end{tabular}
}
\label{tab:lhs}
\caption{Comparing recommendations made by various tuning policies}
\vspace*{-4mm}
\resizebox{\columnwidth}{!}{%
 \begin{tabular}{|c| c c c c c c|}
 \hline
 Application & Policy & \begin{tabular}{@{}c@{}}Containers\\ per Node\end{tabular} & \begin{tabular}{@{}c@{}}Task\\ Concurrency\end{tabular} & \begin{tabular}{@{}c@{}}Cache\\ Capacity\end{tabular} & \begin{tabular}{@{}c@{}}Shuffle\\ Capacity\end{tabular} & \begin{tabular}{@{}c@{}}New\\ Ratio\end{tabular}\\
 \hline
\multirow{3}{*}{\wc} & {\em Exhaustive} & 4 & 2 & 0 & .4 & 1 \\
 & \reinforce & 3 & 2 & 0 & .6 & 3 \\
 & \gaussian & 4 & 2 & 0 & .3 & 1 \\
 & \guided & 4 & 2 & 0 & .3 & 1 \\
 & \relm & 4 & 2 & 0 & .23 & 1 \\
 \hline
\multirow{3}{*}{\sort} & {\em Exhaustive} & 4 & 1 & 0 & .2 & 1 \\
 & \reinforce & 3 & 2 & 0 & .2 & 1 \\
 & \gaussian & 3 & 2 & 0 & .2 & 3 \\
 & \guided & 3 & 2 & 0 & .2 & 1 \\
 & \relm & 4 & 1 & 0 & .23 & 1 \\
 \hline
\multirow{3}{*}{\kmeans} & {\em Exhaustive} & 3 & 2 & .8 & 0 & 7 \\
 & \reinforce & 1 & 4 & .6 & 0 & 4 \\
 & \gaussian & 3 & 1 & .75 & 0 & 3 \\
 & \guided & 3 & 1 & .8 & 0 & 5 \\
 & \relm & 2 & 2 & .68 & 0 & 4 \\
 \hline
\multirow{3}{*}{\svm} & {\em Exhaustive} & 3 & 2 & .8 & .1 & 3 \\
 & \reinforce & 2 & 3 & .6 & .1 & 3 \\
 & \gaussian & 3 & 2 & .2 & .1 & 1 \\
 & \guided & 2 & 3 & .4 & .1 & 3 \\
 & \relm & 3 & 2 & .51 & .07 & 2 \\
 \hline
\multirow{3}{*}{\pagerank} & {\em Exhaustive} & 2 & 1 & .4 & 0 & 3 \\
 & \reinforce & 1 & 4 & .2 & 0 & 5 \\
 & \gaussian & 1 & 2 & .4 & 0 & 3 \\
 & \guided & 2 & 1 & .4 & 0 & 3 \\
 & \relm & 2 & 1 & .24 & 0 & 5 \\
 \hline
 \end{tabular}
}%
\label{tab:main-recos}
\caption{Analysis of a \gaussian\ run for \svm.}
\vspace*{-4mm}
\resizebox{0.8\columnwidth}{!}{%
 \begin{tabular}{|c c c c c | c|}
 \hline
 Sample \# & \begin{tabular}{@{}c@{}}Containers\\ per node\end{tabular} & \begin{tabular}{@{}c@{}}Task\\ concurrency\end{tabular} & \begin{tabular}{@{}c@{}}Cache\\ capacity\end{tabular} & \begin{tabular}{@{}c@{}}New\\ Ratio\end{tabular} & \begin{tabular}{@{}c@{}}Runtime\\ (minutes)\end{tabular}\\
 \hline
 0 & 1 & 4 & 0.6 & 7 & 8.5\\
 0 & 2 & 1 & 0.4 & 3 & 9.3\\
 0 & 3 & 2 & 0.2 & 5 & 7.1\\
 0 & 4 & 2 & 0.8 & 1 & 13\\
 \hline
 1 & 4 & 2 & 0.2 & 5 & 7.3\\
 2 & 2 & 3 & 0.2 & 7 & 7.5\\
 3 & 3 & 2 & 0.2 & 3 & 6.6\\
 4 & 3 & 2 & 0.2 & 1 & {\bf 6.5}\\
 5 & 2 & 3 & 0.2 & 1 & 6.7\\
 6 & 2 & 4 & 0.2 & 1 & 7\\
 \hline
 \end{tabular}
}%
\label{tab:svm-log}
\caption{Comparing tuning algorithm overheads}
\vspace*{-4mm}
\resizebox{0.85\columnwidth}{!}{%
 \begin{tabular}{| l | c | c | c | c |}
 \hline
 Component & \reinforce & \bayesian & \guided & \relm \\
 \hline
Statistics Collection & 5ms & 1ms & 5ms & 5ms \\
Model Fitting & 100ms & 140ms & 180ms & 0.1ms \\
Model Probing & 2ms & 800ms & 1500ms & 0.02ms \\
Model Size & 3Kb & 5Kb & 6Kb & - \\
 \hline
 \end{tabular}
}%
\vspace*{-5mm}
\label{tab:algooverhead}
\end{table}

\begin{figure}
\centering

\begin{tikzpicture}

\begin{axis}[
width=\columnwidth,
height=3.5cm,
ylabel style={align=center}, 
ylabel={Time as \%\\of {\em Exhaustive}\\{\em Search}},
ylabel shift={-4pt},
ybar,
symbolic x coords={1,2,3,4,5},
xticklabels={{\small \wc}, {\small \sort}, {\small \kmeans}, {\small \svm}, {\small \pagerank}},
xtick=data,
bar width=5pt,
enlargelimits=0.1,
every node near coord/.append style={black},
thick,smooth,
]
\addplot+[orange, ybar, postaction={pattern=horizontal lines},
  nodes near coords*={\ifthenelse{\equal{\label}{0}}{}{\label}},
  visualization depends on={value \thisrow{rl-iter} \as \label}
 ]
 table[x=number, y expr = \thisrow{rl-time}*100/\thisrow{exhaustive-time}]
 {pics/training-overhead.tsv};
 
\addplot+[blue, ybar,
  nodes near coords*={\ifthenelse{\equal{\label}{0}}{}{\label}},
  visualization depends on={value \thisrow{gpr-iter} \as \label}
 ]
 table[x=number, y expr = \thisrow{gpr-time}*100/\thisrow{exhaustive-time}]
 {pics/training-overhead.tsv};
 
 \addplot+[magenta, ybar, postaction={pattern=north east lines},
  nodes near coords*={\ifthenelse{\equal{\label}{0}}{}{\label}},
  visualization depends on={value \thisrow{gpropt-iter} \as \label}
 ]
 table[x=number, y expr = \thisrow{gpropt-time}*100/\thisrow{exhaustive-time}]
 {pics/training-overhead.tsv};
 
 \addplot+[green, ybar, postaction={pattern=crosshatch dots},
  nodes near coords*={\ifthenelse{\equal{\label}{0}}{}{\label}},
  visualization depends on={value \thisrow{relm-iter} \as \label}
 ]
 table[x=number, y expr = \thisrow{relm-time}*100/\thisrow{exhaustive-time}]
 {pics/training-overhead.tsv};

\end{axis}

\end{tikzpicture}
\vspace*{-8mm}
\caption{Training overheads of tuning policies. Number of iterations is shown on top of bars.}
 \label{fig:overhead}

\begin{tikzpicture}
\begin{axis}[
width=1.04\columnwidth,
height=4cm,
ylabel={Scaled Runtime},
ylabel shift={-4pt},
ybar,
symbolic x coords={1,2,3,4,5},
xticklabels={{\small \wc}, {\small \sort}, {\small \kmeans}, {\small \svm}, {\small \pagerank}},
xtick=data,
bar width=4pt,
enlargelimits=0.1,
legend columns=3,
legend style={at={(0.5,1.5)},anchor=north},
every node near coord/.append style={black},
legend cell align={left},
thick,smooth,
]
 \addplot+[red, ybar, postaction={pattern=north west lines},
  nodes near coords*={\ifthenelse{\equal{\label}{0}}{}{\label}},
  visualization depends on={value \thisrow{failed1} \as \label}
 ] table[x=App, y=sRT1]{pics/main-results.dat};
 \addlegendentry{\maximizeresource};
  \addplot+[orange, ybar, postaction={pattern=horizontal lines},
  nodes near coords*={\ifthenelse{\equal{\label}{0}}{}{\label}},
  visualization depends on={value \thisrow{failed6} \as \label}
 ] table[x=App, y=sRT6]{pics/main-results.dat};
 \addlegendentry{\reinforce};
\addplot+[yellow, ybar, postaction={pattern=grid},
  nodes near coords*={\ifthenelse{\equal{\label}{0}}{}{\label}},
  visualization depends on={value \thisrow{failed2} \as \label}
 ] table[x=App, y=sRT2]{pics/main-results.dat};
 \addlegendentry{\em Exhaustive};
\addplot+[blue, ybar,
  nodes near coords*={\ifthenelse{\equal{\label}{0}}{}{\label}},
  visualization depends on={value \thisrow{failed3} \as \label}
 ] table[x=App, y=sRT3,
 ]{pics/main-results.dat};
 \addlegendentry{\bayesian};
 \addplot+[magenta, ybar, postaction={pattern=north east lines},
  nodes near coords*={\ifthenelse{\equal{\label}{0}}{}{\label}},
  visualization depends on={value \thisrow{failed4} \as \label}
 ] table[x=App, y=sRT4,
 ]{pics/main-results.dat};
 \addlegendentry{\guided};
 \addplot+[green, ybar, postaction={pattern=crosshatch dots}] table[x=App, y=sRT5]{pics/main-results.dat};
 \addlegendentry{\relm};
\end{axis}

\end{tikzpicture}
\vspace*{-6mm}
\caption{Runtime of every recommended configuration is scaled to the runtime of \maximizeresource. Number of failed containers is shown on top of bars.}
\label{fig:main}
\vspace*{-7mm}
\end{figure}

\begin{figure}
\centering

\begin{tikzpicture}
\begin{axis}[
width=0.7\columnwidth,
height=3.5cm,
xlabel={Runtime(min)},
ytick={1,2},
yticklabels={{\bayesian}, {\guided}},
/pgfplots/boxplot/box extend=0.4,
enlargelimits=0.25,
thick,smooth,
]
    [above]
node at
(boxplot box cs: \boxplotvalue{lower quartile},1)
{\pgfmathprintnumber{0}}
  ;

 \addplot+[blue, boxplot, boxplot/draw position=1]
  table [row sep=\\,y index=0] {
  	data\\
	106\\ 107\\ 287\\ 98\\ 226\\ 214\\ 58\\ 205\\ 309\\ 123\\
  }
  [above]
node at
(boxplot box cs: \boxplotvalue{lower quartile},1)
{\pgfmathprintnumber{7}}
node[left] at
(boxplot box cs: \boxplotvalue{median},0.5)
{\pgfmathprintnumber{10}}
node at
(boxplot box cs: \boxplotvalue{upper quartile},1)
{\pgfmathprintnumber{13}}
  ;

 \addplot+[magenta, boxplot, boxplot/draw position=2]
  table [row sep=\\,y index=0] {
  	data\\
	48\\ 112\\ 94\\ 90\\ 63\\
  }
  [above]
 node at
(boxplot box cs: \boxplotvalue{lower quartile},1)
{\pgfmathprintnumber{4}}
node[left] at
(boxplot box cs: \boxplotvalue{median},0.5)
{\pgfmathprintnumber{5}}
node at
(boxplot box cs: \boxplotvalue{upper quartile},1)
{\pgfmathprintnumber{6}} 
 ;
\end{axis}

\end{tikzpicture}
\vspace*{-5mm}
\caption{Training time and the number of iterations for \kmeans\ after executing each policy 5-10 times}
\label{fig:kmeans-box}

\vspace*{4mm}
\begin{tikzpicture}
\begin{axis}[
width=0.7\columnwidth,
height=3.5cm,
xlabel={Runtime(min)},
ytick={1,2},
yticklabels={{\bayesian}, {\guided}},
/pgfplots/boxplot/box extend=0.4,
enlargelimits=0.25,
thick,smooth,
]
    [above]
node at
(boxplot box cs: \boxplotvalue{lower quartile},1)
{\pgfmathprintnumber{0}}
  ;

 \addplot+[blue, boxplot, boxplot/draw position=1]
  table [row sep=\\,y index=0] {
  	data\\
	28.9\\ 29.9\\ 32.4\\ 39.5\\ 46.6\\ 63.2\\ 69.9\\ 83.7\\ 128.5\\ 145.5\\
  }
  [above]
node at
(boxplot box cs: \boxplotvalue{lower quartile},1)
{\pgfmathprintnumber{5}}
node[left] at
(boxplot box cs: \boxplotvalue{median},0.5)
{\pgfmathprintnumber{9}}
node at
(boxplot box cs: \boxplotvalue{upper quartile},1)
{\pgfmathprintnumber{13}}
  ;

 \addplot+[magenta, boxplot, boxplot/draw position=2]
  table [row sep=\\,y index=0] {
  	data\\
	4.4\\ 5.2\\ 16.4\\ 29.1\\ 42.4\\ 44.1\\ 64.1\\ 64.3\\ 135.3\\
  }
  [above]
 node at
(boxplot box cs: \boxplotvalue{lower quartile},1)
{\pgfmathprintnumber{3}}
node[left] at
(boxplot box cs: \boxplotvalue{median},0.5)
{\pgfmathprintnumber{6}}
node at
(boxplot box cs: \boxplotvalue{upper quartile},1)
{\pgfmathprintnumber{10}} 
 ;
\end{axis}

\end{tikzpicture}
\vspace*{-5mm}
\caption{Training time and the number of iterations for \svm\ after executing each policy 5-10 times}
\label{fig:svm-box}
\end{figure}

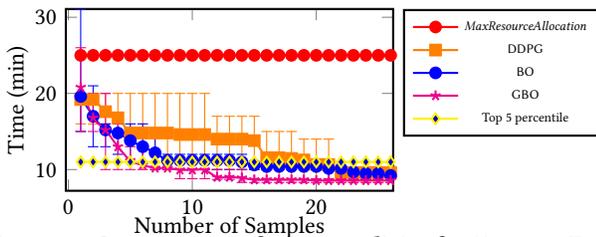
\begin{figure}
\centering
\begin{tikzpicture}
\begin{axis}[
width=0.7\columnwidth,
height=4cm,
ylabel style={align=center},
ylabel={Time (min)},
ylabel shift={-4pt},
ymax=30,
xmax=25,
xlabel={Number of Samples},
xlabel style={yshift=1.5ex},
enlargelimits=0.05,
thick,smooth,
legend pos=outer north east,
legend style={font=\tiny},
]
 \addplot+[red, mark options={fill=red}] 
  table[x=iteration, y expr=25]{pics/kmeans-convergence.tsv};
  \addlegendentry{\maximizeresource}

 \addplot+[orange, mark options={fill=orange}, error bars/.cd, y fixed, y dir=both, y explicit] 
  table[x=iteration, y=rl-mean, y error plus expr=\thisrow{rl-max}-\thisrow{rl-mean}, y error minus expr=\thisrow{rl-mean}-\thisrow{rl-min}]{pics/kmeans-convergence.tsv};
  \addlegendentry{\reinforce}

 \addplot+[blue, mark options={fill=blue}, error bars/.cd, y fixed, y dir=both, y explicit] 
  table[x=iteration, y=gprei-mean, y error plus expr=\thisrow{gprei-max}-\thisrow{gprei-mean}, y error minus expr=\thisrow{gprei-mean}-\thisrow{gprei-min}]{pics/kmeans-convergence.tsv};
  \addlegendentry{\bayesian}

 \addplot+[magenta, mark options={fill=magenta}, error bars/.cd, y fixed, y dir=both, y explicit] 
  table[x=iteration, y=gpreiopt-mean, y error plus expr=\thisrow{gpreiopt-max}-\thisrow{gpreiopt-mean}, y error minus expr=\thisrow{gpreiopt-mean}-\thisrow{gpreiopt-min}]{pics/kmeans-convergence.tsv};
  \addlegendentry{\guided}

 \addplot+[yellow] 
 table[x=iteration, y expr=11]{pics/kmeans-convergence.tsv};
 \addlegendentry{Top 5 percentile}

\end{axis}
\end{tikzpicture}
\vspace*{-6mm}
\caption{Convergence of tuning policies for \kmeans. Each tuner is run 5 times; the mean, min, and max values for the lowest runtime observed so far are plotted.}
 \label{fig:convergence}
\vspace*{-7mm}
\end{figure}

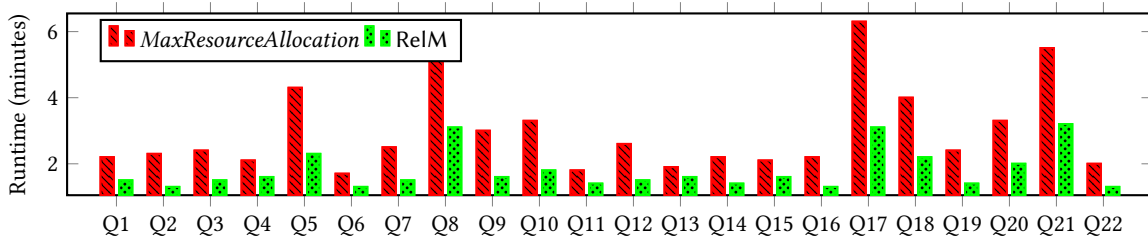
\begin{figure*}
\centering
\begin{tikzpicture}
\begin{axis}[
width=0.9\textwidth,
height=4cm,
ylabel={Runtime (minutes)},
ybar,
symbolic x coords={1,2,3,4,5,6,7,8,9,10,11,12,13,14,15,16,17,18,19,20,21,22},
xticklabels={Q1,Q2,Q3,Q4,Q5,Q6,Q7,Q8,Q9,Q10,Q11,Q12,Q13,Q14,Q15,Q16,Q17,Q18,Q19,Q20,Q21,Q22},
xtick=data,
bar width=5pt,
enlargelimits=0.05,
legend columns=-1,
legend pos=north west,
legend entries={\maximizeresource, \relm},
thick,smooth,
]
 \addplot+[red, ybar, postaction={pattern=north west lines}] table[x=Query, y=Runtime1]{pics/tpch-results.dat};

 \addplot+[green, ybar, postaction={pattern=crosshatch dots}] table[x=Query, y=Runtime2]{pics/tpch-results.dat};
 
\end{axis}

\end{tikzpicture}
\vspace*{-5mm}
\caption{Performance comparison of TPC-H Queries run using \maximizeresource\ policy and using \relm.}
 \label{fig:tpch}
\vspace*{-5mm}
\end{figure*}

The first question we want to answer is \textbf{\textit{how long does it take to produce high quality tuning results?}}. We carry out \exhaustive\ on Cluster A and use it as a baseline for other policies. The black-box policies are trained on each application individually until they find a configuration with performance within top 5 percentile of the baseline. The process is repeated 5 to 10 times and only the mean values of overheads are plotted in Figure~\ref{fig:overhead}.

\relm\ needs a single application run in each case to analytically find a desired configuration. So it has the lowest overhead. The regression policies, \bayesian\ and \guided, require less than 4\% of the effort needed for \exhaustive\ with \guided\ being about 2 times faster. The \reinforce\ policy takes longer, but still less than 10\% time compared to the exhaustive search.

Figure~\ref{fig:overhead} compared the training times against a baseline of \exhaustive. In order to get a sense of the absolute times required and also to understand the variability of results across multiple runs, we include a couple of box-whisker plots in Figure~\ref{fig:kmeans-box} and Figure~\ref{fig:svm-box} respectively. The lower and the upper quantiles represent the 25th and the 50th percentile respectively, while the vertical line within the box represents the median runtime. The numbers provided along with the box denote the number of training iterations corresponding to the respective quantiles.

We do notice considerable variations in terms of the training overheads across runs. A major reason behind this is the phenomenon of local minima observed by the black-box model. It is particularly noticeable in both the policies on \svm\ application where the distribution has a long tail. \reinforce\ suffers from the same issue and is seen to take the longest amongst the black-box policies to optimize. 

\noindent \textbf {\textit{Black-box models can get stuck in a local minima.}}\\
Figure~\ref{fig:convergence} shows an example run showing how the policies converge. For the first 12 iterations, \reinforce\ tries out configurations with lower values for \poolcache\ with very low rewards. Post which, it starts exploring higher values for cache, higher rewards follow, and the model converges to the desired performance. Between \bayesian\ and \guided, we observe that \guided\ model fits data earlier compared to \bayesian. Section~\ref{sec:bo-eval} carries out an analysis using a validation set which corroborates the observation.

It is also found that the quality of results of \gaussian, to a large extent, depends on the initial samples used in bootstrapping. We provide a log of \gaussian\ run for \svm\ in Table~\ref{tab:svm-log}. Based on the initial samples, \gaussian\ pins down the Cache Capacity to 0.2 and continues exploring the other parameters. The application requires a capacity over 0.5 to fit in the entire cached data in memory. While this fact is captured in the white-box models of \relm, \gaussian\ fails to explore this region. \guided, though not exempt, tends to come out of a local minima quicker because of the additional features from model $Q$ guiding the exploration.

The second question we want to analyze is \textbf{\textit{how much performance improvement is exhibited by our tuning policies?}}. We use a stopping criteria for black-box exploration policies: Bayesian policies are executed until the expected improvement falls below 10\% and at least 6 new samples have been observed in addition to the 4 LHS samples~\cite{cherrypick}; \reinforce\ is similarly stopped when it has observed 10 new samples. Although both the policies are capable of re-using models from prior tuning runs, we train them with a cold start in this evaluation; model re-use is discussed in Section~\ref{sec:reuse}.
 
Figure~\ref{fig:main} compares the performance. The runtime of every recommended configuration is scaled to the runtime of the default policy for the same application. The number of failed containers is indicated by a label over the corresponding bar. \relm\ consistently achieves a runtime within 10\% of the best configuration found using \exhaustive. Moreover, \relm\ ensures no containers run out of memory. 
Table~\ref{tab:main-recos} lists the recommendations made by tuning policies. In the case of \kmeans, it can be noted that a high memory fraction is allocated to \poolcache\ in the configurations found by the policies of {\em Exhaustive}, \bayesian, and \guided\ leaving very small memory for other objects. 
Similar observations can be made about the container failures exhibited by other policies as well. 
\relm\ avoids this issue by treating safety as the first class citizen in its modeling.

The performance improvement over the default setup, in most cases, is between 50\%-70\%. 
In the case of \svm, however, \bayesian\ and \guided\ policies find configurations that are better than the default ones by only 10\% and 20\% respectively. This happens due to exploration hitting a local minima.

In addition to the possibility of getting stuck in a local minima, another concern with the black-box models is setting the right objective function. With the objective set to minimize runtime in our evaluation, \guided\ recommends a configuration for \pagerank\ which is {\em unreliable}. As a  workaround, the \gaussian\ algorithm should be given an objective function that incorporates penalties for such failures. It is, however, hard for users to find the right objective function. \relm, on the other hand, has safety built into its model and therefore, tends to recommend only {\em safe} configurations.

\vspace*{-2mm}
\subsection{Algorithm Overheads}
\label{sec:overheads}
\vspace*{-1mm}

Overheads presented in Figure~\ref{fig:overhead} are largely dominated by observation (stress testing) times. We focus on the other components here, viz. (1) Statistics collection, (2) Model fitting, and (3) Model probe. Table~\ref{tab:algooverhead} compares one iteration from each algorithm. Except \bayesian, all algorithms involve collecting internal resource usage statistics to build either white-box models or state metrics. 

While model fitting involves an update of the actor-critic networks in \reinforce, it requires an update of Gaussian Process with a new observation in \bayesian. The higher overhead for \guided\ compared to \bayesian\ is down to the added dimensionality due to model $Q$. The same is true when probing the model which involves computing expected improvement on a sample of configuration space. These numbers show why the \bayesian\ regression model is not suited for high dimensional spaces. 

In the case of \relm, both model fitting, which evaluates a small series of analytical models, and model probe, which involves looping through a small handful of container configurations, are inexpensive. We performed a small scalability test by artificially creating 100 container configurations, a considerably large number compared to the practical cluster setups. The model probe time goes up to 10ms which, though is a considerable increase from our test environment, is a small overhead when compared to other algorithms.

 The black-box models can be saved for later use if an application similar to previously seen one is to be tuned. We compare the storage overhead of the models for the same. While \reinforce\ stores the learned parameters of the actor-critic neural network, \bayesian\ stores the entire training data. Though the last row of Table~\ref{tab:algooverhead} shows that both \reinforce\ and \bayesian\ have a small storage overhead, the size of \bayesian\ model grows linearly with training data making it suitable only when the number of samples used for training is small.

\vspace*{-2mm}
\subsection{Analysis of \relm}
\label{sec:relm-eval}
\vspace*{-1mm}

\noindent \textbf {\textit{\relm\ is robust to workload variations.}}\\
\relm\ does not depend on application workflow and input data design directly, but rather uses interactions between configuration options and resource metrics in its models. It, therefore, can handle different workload types as evident from Figure~\ref{fig:main}. 
We also evaluate TPC-H  benchmark workload on Cluster B to further press the point. As seen from Figure~\ref{fig:tpch}, the workload when executed using \maximizeresource\ takes a total of 66 minutes. Using profile of this run, \relm\ cuts the runtime down to 40 minutes, a saving of 40\%.

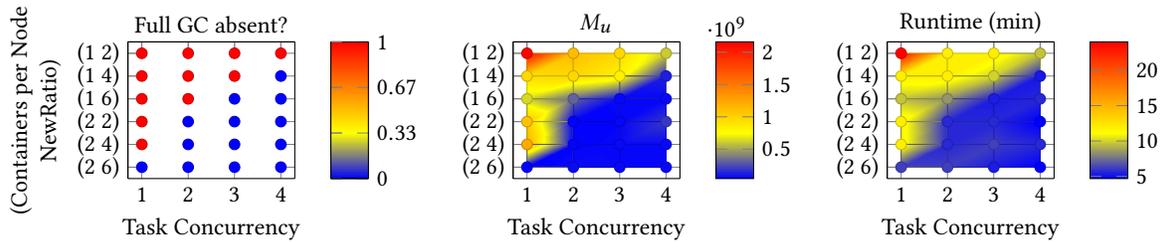
\begin{figure*}
\centering
\begin{tikzpicture}
\begin{axis}[
width=3.8cm,
height=3.4cm,
title={Full GC absent?},
title style={yshift=-1.5ex},
xlabel={Task Concurrency},
ylabel style={align=center},
ylabel={(Containers per Node\\ NewRatio)},
view={0}{90},
xtick=data, 
ytick=data,
yticklabels={(1 2), (1 4), (1 6), (2 2), (2 4), (2 6)},
colorbar right,
colorbar style={ytick=data,},
y dir=reverse,
enlargelimits=0.1,
]
\addplot3[shader=flat, scatter, only marks, mark=*, mesh/rows=4] coordinates
{
 (1, 1, 1) (1, 2, 1) (1, 3, 1) (1, 4, 1) (1, 5, 1) (1, 6, 0)
 (2, 1, 1) (2, 2, 1) (2, 3, 1) (2, 4, 0) (2, 5, 0) (2, 6, 0)
 (3, 1, 1) (3, 2, 1) (3, 3, 0) (3, 4, 0) (3, 5, 0) (3, 6, 0)
 (4, 1, 1) (4, 2, 0) (4, 3, 0) (4, 4, 0) (4, 5, 0) (4, 6, 0)
};
\end{axis}
\end{tikzpicture}
\quad
\begin{tikzpicture}
\begin{axis}[
height=3.4cm,
width=3.8cm,
title={$M_{u}$},
title style={yshift=-1.5ex},
xlabel={Task Concurrency},
view={0}{90},
xtick=data,
ytick=data, 
yticklabels={(1 2), (1 4), (1 6), (2 2), (2 4), (2 6)},
colorbar,
y dir=reverse,
enlargelimits=0.1,
]
\addplot3[surf, shader=faceted interp, scatter, mark=*, mesh/rows=4] coordinates
{
 (1, 1, 2.16E+09) (1, 2, 9.59E+08) (1, 3, 6.37E+08) (1, 4, 1.09E+09) (1, 5, 1.22E+09) (1, 6, 4.47E+07)
 (2, 1, 1.12E+09) (2, 2, 9.69E+08) (2, 3, 3.19E+08) (2, 4, 5.21E+07) (2, 5, 5.53E+07) (2, 6, 7.61E+07)
 (3, 1, 9.69E+08) (3, 2, 8.34E+08) (3, 3, 6.43E+07) (3, 4, 6.05E+07) (3, 5, 6.60E+07) (3, 6, 6.84E+07)
 (4, 1, 6.02E+08) (4, 2, 1.28E+08) (4, 3, 6.81E+07) (4, 4, 1.95E+08) (4, 5, 6.97E+07) (4, 6, 6.66E+07)
};
\end{axis}
\end{tikzpicture}
\quad
\begin{tikzpicture}
\begin{axis}[
width=3.8cm,
height=3.4cm,
title={Runtime (min)},
title style={yshift=-1.5ex},
xlabel={Task Concurrency},
view={0}{90},
xtick=data, 
ytick=data,
yticklabels={(1 2), (1 4), (1 6), (2 2), (2 4), (2 6)},
colorbar,
y dir=reverse,
enlargelimits=0.1,
]
\addplot3[surf, shader=faceted interp, scatter, mark=*, mesh/rows=4] coordinates
{
 (1, 1, 24) (1, 2, 12) (1, 3, 9.6) (1, 4, 12) (1, 5, 12) (1, 6, 6.3)
 (2, 1, 12) (2, 2, 12) (2, 3, 8.4) (2, 4, 5.7) (2, 5, 5.7) (2, 6, 6.3)
 (3, 1, 12) (3, 2, 10) (3, 3, 5.4) (3, 4, 5.7) (3, 5, 6.3) (3, 6, 5.4)
 (4, 1, 9.6) (4, 2, 5.3) (4, 3, 5.7) (4, 4, 4.7) (4, 5, 5.7) (4, 6, 4.7)
};
\end{axis}
\end{tikzpicture}
\vspace*{-0.5cm}
\caption{Understanding sensitivity of \relm\ recommendations to the initial profile. When no full GC events are present, maximum \poolold\ pool value is used to estimate task memory. This results in an over-estimate of memory requirements and sub-optimal recommendations. Profiles with full GC events, on the other hand, produce more accurate estimates.}
\vspace*{-6mm}
 \label{fig:svm}
\end{figure*}

\begin{figure}
\centering
\begin{tikzpicture}

\begin{axis}[
width=\columnwidth,
height=3.6cm,
ylabel={Memory in Bytes},
symbolic x coords={1,2,3,4,5},
xticklabels={{\wc}, {\sort}, {\kmeans}, { \svm}, {\pagerank}},
xtick=data,
legend pos=north west,
legend columns=-1,
ymode=log,
smooth, thick,
enlargelimits=0.1,
ybar,
nodes near coords={
        },
every node near coord/.append style={font=\footnotesize},
]
\pgfplotsset{cycle list shift=1}
\addplot+[ybar, error bars/.cd, y dir=both, y explicit] 
table[x=App, y=Init, y error=InitDev]{pics/training-accuracy.dat};
\addlegendentry{$M_{i}$}
\pgfplotsset{cycle list shift=2}
\addplot+[ybar, postaction={pattern=north west lines}, error bars/.cd, y dir=both, y explicit] 
table[x=App, y=Task, y error=TaskDev]{pics/training-accuracy.dat};
\addlegendentry{$M_{u}$}
\end{axis}
\end{tikzpicture}
\vspace*{-0.5cm}
\caption{Analyzing sensitivity to initial profile by invoking \relm\ with 16 unique initial profiles. Error bars indicate the standard error of the mean.}
 \label{fig:sense}
\vspace*{-7mm}
\end{figure}
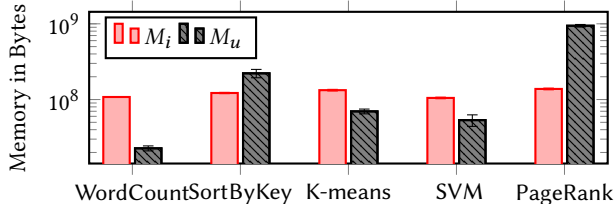

\noindent \textbf{\textit{\relm\ requires full GC events in input profile.}}\\
We use the \svm\ application to study \relm's sensitivity to the initial profile. \svm\ tasks use a small amount of memory because of the small-sized partitions (see Table~\ref{tab:benchmarks}). For sufficiently large \poolheap\ values, this memory can be collected by young GC events with very few objects tenuring to \poolold. Further, the cache requirement of only about 50\% \poolheap\ creates very low pressure on \poolold\ pool. We exhaustively invoked \relm\ with profiles generated from multiple configurations of \svm\ on our cluster and analyzed the recommendations. 

Figure~\ref{fig:svm} shows that the memory requirements are over-estimated by up to 2 orders of magnitude when using the profiles containing no full GC events. \relm\ tackles such cases by making an additional profiling run with its configuration options set using simple heuristics discussed in Section~\ref{sec:profiling}.

\noindent \textbf{\textit{\relm\ is robust to input profiles containing full GC events.}}\\
It can be observed from Figure~\ref{fig:svm} that the runtimes exhibited by the recommendations made from the input profiles containing full GC events are clustered around 5 minutes. In fact, there are only 3 unique recommendations in the bottom right quadrant differing very slightly 
in terms of the Task Concurrency and Cache Capacity values. 
We include a plot in Figure~\ref{fig:sense} showing the variability in the estimates of the \poolcode\ ($M_{i}$) and \poolunmgd\ ($M_{u})$ pools by using different profiles containing full GC events. It can be noticed that the estimated memory requirements have very little variance. The algorithm, as a result, recommends the same (with minor changes) configuration no matter where we start. We have used logarithmic scale for better visibility since the task memory values across the applications differ by up to two orders of magnitude.

\begin{figure}
\centering
\begin{tikzpicture}
\begin{axis}[
width=0.45\columnwidth,
title={\rmv},
title style={yshift=-1.5ex},
ylabel={\#Containers},
view={0}{90},
symbolic x coords={1,2,3,4,5},
xticklabels={{\tiny \wc}, {\tiny \sort}, {\tiny \kmeans}, {\tiny \svm}, {\tiny \pagerank}},
xtick=data,
ytick=data,
x tick label style={rotate=30,anchor=east},
colorbar horizontal,
colorbar style={at={(0.5, -0.5)},anchor=south, height=3mm},
colormap/cool,
visualization depends on={\thisrow{RMVRank} \as \rnk},
scatter/@pre marker code/.append style=
{/tikz/mark size={1pt+abs(4-\rnk)}},
enlargelimits=0.15,
]
\addplot3[shader=flat, scatter, only marks,] 
table[x=App, y=NumExecs, z=RMV] {pics/relm-accuracy.dat};
\end{axis}
\end{tikzpicture}
\quad
\begin{tikzpicture}
\begin{axis}[
width=0.45\columnwidth,
title={Scaled Runtime},
title style={yshift=-1.5ex},
ylabel={\#Containers},
view={0}{90},
symbolic x coords={1,2,3,4,5},
xticklabels={{\tiny \wc}, {\tiny \sort}, {\tiny \kmeans}, {\tiny \svm}, {\tiny \pagerank}},
xtick=data,
ytick=data,
x tick label style={rotate=30,anchor=east},
colorbar horizontal,
colorbar style={at={(0.5, -0.5)},anchor=south, height=3mm},
colormap/greenyellow,
visualization depends on={\thisrow{RuntimeRank} \as \rnk},
scatter/@pre marker code/.append style=
{/tikz/mark size={.8pt+abs(4-\rnk)}},
enlargelimits=0.15,
]
\addplot3[shader=flat, scatter, only marks,] 
table[x=App, y=NumExecs, z=Runtime] {pics/relm-accuracy.dat};
\end{axis}
\end{tikzpicture}
\vspace*{-5mm}
\caption{Evaluating accuracy of configurations ranking in \relm. Bubble size indicates the rank of the point in its column. Bigger bubble signifies a high \rmv\ score/ a low runtime.}
\vspace*{-5mm}
 \label{fig:accuracy}
\end{figure}
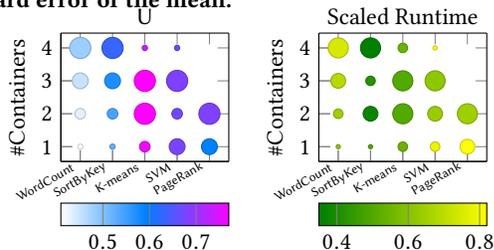

\noindent \textbf{\textit{\relm\ correctly ranks the candidate configurations.}}\\
\relm\ ranks the best configuration found on every enumerated candidate container size by the utility score \rmv. Figure~\ref{fig:accuracy} compares the ranks of the configurations based on \rmv\ to those based on their performance and finds a strong correlation between the two.
\eat{
\relm\ enumerates each candidate container size and ranks the recommendations using the utility score \rmv. The utility score corresponds to the total memory utilization of the internal pools combined. It should be noted that the utility score does not correspond to the expected \poolheap\ utilization because the formulation does not factor in garbage objects in \poolheap. The expectation, however, is that configuration with a higher utility score will perform better because of the higher memory provisioned for internal pools.

We verify our intuition by plotting the \rmv\ scores and the runtimes for the configurations considered while tuning our test applications. Figure~\ref{fig:accuracy} shows the results. The bubble size for a configuration is set to denote its rank relative to  the other configurations considered for the application under test. 
The ranks obtained based on runtimes indicate a direct correlation to the ranks based on \rmv\ values. The instances of inversions are the cases where the values are very close to each other (indicated by color legends) and do not affect the quality of results significantly. In practice, the end user could be presented multiple configuration options to choose from all having \rmv\ values close to the best.
}

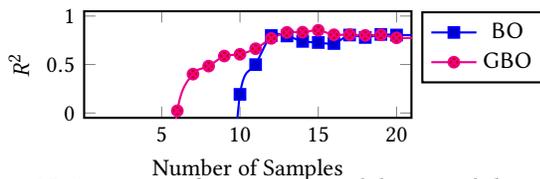
\begin{figure}
\centering

 \begin{tikzpicture}
\begin{axis}[
width=0.7\columnwidth,
height=3cm,
ylabel style={align=center},
ylabel={$R^2$},
ymin=0, ymax=1,
xmin=1, xmax=20,
xlabel={Number of Samples},
enlargelimits=0.05,
thick,smooth,
legend pos=outer north east,
]
\addplot+[blue, mark=square*] 
 table[x=iteration, y=gpr-r2]{pics/kmeans-convergence.tsv};
 \addlegendentry{\bayesian}

\addplot+[magenta, mark=otimes*] 
 table[x=iteration, y=gpropt-r2]{pics/kmeans-convergence.tsv};
 \addlegendentry{\guided}

\end{axis}
\end{tikzpicture}
\vspace*{-6mm}
\caption{Accuracy of surrogate model on a validation set. Higher values indicate better fit.}
 \label{fig:r2}
 
\vspace*{-6mm}
\end{figure}

\vspace*{-3mm}
\subsection{Analysis of \guided}
\label{sec:bo-eval}
\vspace*{-1mm}


In order to understand the speedup in \guided\ over \bayesian, we study the quality of the models on a validation set which corresponds to about 10\% of the configurations considered by \exhaustive. Figure~\ref{fig:r2} plots the Coefficient of Determination~\cite{r2} on the validation set observed after each iteration. Accuracy of the \bayesian\ model is very poor until 10 iterations, while \guided\ starts fitting decent models much earlier. This improvement can be attributed to the white-box features added by \guided: We analyzed the correlation of each individual feature to the performance objective using Pearson Correlation Coefficient~\cite{pearson}. It is found that the feature that shows the highest correlation in \bayesian\ corresponds to the Cache Capacity setting. The model developed by \guided\ using the same number of samples, on the other hand, shows that the two of the three newly added features by model $Q$, namely $q_1$ and $q_2$ from Eq.~\ref{eq:q}, show an even stronger correlation. 

\begin{figure}
\centering
\begin{tikzpicture}
\begin{axis}[
width=\columnwidth,
height=4.2cm,
ylabel={Training Time (min)},
ybar,
symbolic x coords={1,2},
xticklabels={{\kmeans}, {\svm}},
xtick=data,
enlarge y limits=0.2,
enlarge x limits=0.4,
legend columns=2,
legend pos=north east,
every node near coord/.append style={black},
legend cell align={left},
thick,smooth,
]
\addplot+[blue, ybar,
  nodes near coords*={\ifthenelse{\equal{\label}{0}}{}{\label}},
  visualization depends on={value \thisrow{iter-bo-gpr} \as \label}
 ] table[x=App, y=time-bo-gpr, 
 ]{pics/surrogate.txt};
 \addlegendentry{\bayesian-GP};
 \addplot+[magenta, ybar, postaction={pattern=north east lines},
  nodes near coords*={\ifthenelse{\equal{\label}{0}}{}{\label}},
  visualization depends on={value \thisrow{iter-gbo-gpr} \as \label}
 ] table[x=App, y=time-gbo-gpr,
 ]{pics/surrogate.txt};
 \addlegendentry{\guided-GP};
\addplot+[cyan, ybar,
  nodes near coords*={\ifthenelse{\equal{\label}{0}}{}{\label}},
  visualization depends on={value \thisrow{iter-bo-rf} \as \label}
 ] table[x=App, y=time-bo-rf, 
 ]{pics/surrogate.txt};
 \addlegendentry{\bayesian-RF};
 \addplot+[orange, ybar, postaction={pattern=north east lines},
  nodes near coords*={\ifthenelse{\equal{\label}{0}}{}{\label}},
  visualization depends on={value \thisrow{iter-gbo-rf} \as \label}
 ] table[x=App, y=time-gbo-rf,
 ]{pics/surrogate.txt};
 \addlegendentry{\guided-RF};
\end{axis}

\end{tikzpicture}
\vspace*{-5mm}
\caption{Comparing the impact of changing the surrogate model from Gaussian Process (GP) to Random Forest (RF). Numbers on top of the bars denote the number of iterations.}
\label{fig:rf}
\vspace*{-5mm}
\end{figure}
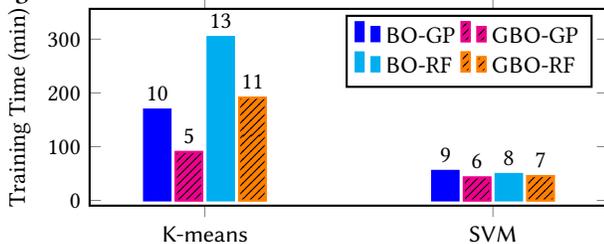

We have used Gaussian Process (GP) as a surrogate model both to illustrate the \bayesian\ mechanism as well as during the evaluation because of its salient features such as confidence bound on predictions, support for noisy observations, and an ability to use gradient-based methods~\cite{ieee16}. Alternate ensemble tree-based models such as Random Forest and Boosted Regression Trees have been shown to be better at modeling the non-linear interactions~\cite{arrow}. However, they lack theoretical guarantees on the confidence bounds that the Gaussian Process offers. We do a small experiment to analyze if an alternate model would be a better fit for our scenario.

We evaluate Random Forest (RF) regression model to tune \kmeans\ and \svm\ applications. The results are compared against the Gaussian Process (GP) in Figure~\ref{fig:rf}. It is not clear from the graph if one model is strictly superior over the other: e.g. GP fits \kmeans\ application better, whereas RF is able to explain the interactions among the configuration options in \svm\ better. However, the \guided\ framework helps no matter which surrogate model is used under the covers. 

It should be pointed that a more extensive analysis of the impact of the surrogate models and the hyperparameters defining them is left as a future work. While the hyperparameter choices in case of the Gaussian Process include the choice of the kernel and the parameters thereof used as a covariance metrics~\cite{ieee16}, the hyperparameters for tree-based ensemble models include the parameters relating to the dimensions of the trees~\cite{forest}. A detailed impact analysis is left out of scope for this paper.

\vspace*{-3mm}
\subsection{Generality of models}
\label{sec:reuse}
\vspace*{-1mm}

\begin{figure}
\begin{tikzpicture}
\begin{axis}[
ybar,
width=0.9\linewidth,
height=0.4\linewidth,
ylabel={Runtime (min)},
symbolic x coords={1, 2, 3, 4},
xticklabels={{$\reinforce_{A}^{B}$}, {$\reinforce_{B}^{B}$}, {$\reinforce_{s1}^{s2}$}, {$\reinforce_{s2}^{s2}$}},
xtick=data,
ymin=1,
smooth, thick,
enlargelimits=0.15,
]
\addplot+[ybar, error bars/.cd, y dir=both, y explicit] coordinates
{
 (1, 4.02) += (0, .5) -= (0, .2)
 (2, 3.96) += (0, .1) -= (0, .16)
 (3, 9.76) += (0, 2.24) -= (0, 1.46)
 (4, 9.46) += (0, 1.54) -= (0, 1.16)
 };
\end{axis}
\end{tikzpicture}
\vspace*{-0.5cm}
\caption{Studying generality of \reinforce\ by applying it to a different cluster and to a different input dataset}
\vspace*{-0.7cm}
 \label{fig:reuse}
\end{figure}
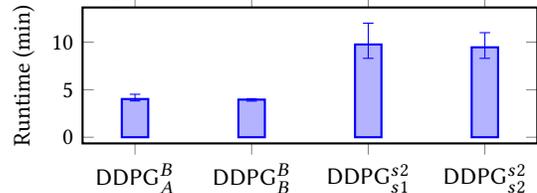

We analyze how our tuning policies adapt to a new environment or a new workload. As \relm\ takes a profile-based white-box approach to tuning, it needs at least a single run in the test environment. We have shown that a single profiling run is often sufficient as well since it contains enough information of the expected memory usage of both the resource containers and internal memory pools. Adaptability of \relm\ is evident from tests carried out on varied computational patterns, data layout (partition size), and   resource clusters.

Black-box tuning policies, however, need to find ways to generalize models in order to reduce the stress testing time. OtterTune~\cite{ottertune} re-uses Bayesian model trained on a prior workload by mapping the present workload based on the measurements of a set of external performance metrics. The OtterTune strategy is replicated in our setup by matching two applications based on the performance statistics (shown in Table~\ref{tab:stats}) derived on the default configuration. However, the saved regression models cannot be adapted to the changes in hardware configuration and input data.

Unlike the performance-based regression model of \bayesian, the \reinforce\ model is trained using reward-feedback. It, therefore, showcases better adaptability to changes in test environment. We present an evaluation in Figure~\ref{fig:reuse}. First, we use a model trained on Cluster A to cross test the same workload, SVM application in this case, on Cluster B (denoted by $\reinforce_{A}^{B}$). Its output is compared with the output produced by a model trained on Cluster B ($\reinforce_{B}^{B}$). The cross testing is allowed to use only 5 test samples. By using the insights gained during prior training, the \reinforce\ policy can quickly adapt to the hardware changes. Another experiment carried out by changing the input data scale factor for SVM workload on Cluster B (from s1 to s2) shows similar observation.

\vspace*{-3mm}
\subsection*{Summary}
\vspace*{-1mm}
The tuning policies we evaluated each have their strong points. The evaluation justifies our approach of modeling interactions between the memory configuration options using which \relm\ model provides a good recommendation very quickly. Bayesian regression policies can provide optimality guarantees at a higher training cost; with \guided\ speeding up the exploration by 2x. \reinforce\ policy supports an equally powerful AI-based tuning with minimal algorithm overheads and better adaptability to changes in environment making it an attractive choice for tuning problems where no simple white-box models can assist.

\vspace*{-3mm}
\section{Related Work}
\label{sec:related}
\vspace*{-1mm}

There has been a large body of work on auto-tuning the physical design of database systems~\cite{physical} which includes index selection~\cite{index}, data partitioning~\cite{partitioning}, and view materialization~\cite{materialization}. Comparatively less work has looked at on auto-tuning the internal configuration parameters like memory pools. Most commercial database systems provide configuration tuning wizards to DBAs which, based on the user feedback on workload performance, suggest better settings for configuration parameters using white-box models~\cite{db2advisor}. DB2 provides a Self-tuning Memory Manager (STMM)~\cite{stmm} which uses analytical models to determine cost-benefits of the internal memory pools. Oracle's {\em ADDM} can identify performance bottlenecks due to misconfigurations and recommend necessary changes \cite{addm}. 

Recent attempts at auto-tuning systems have either focussed on building {\em What-If} performance models~\cite{starfish, mrtuner, tempo} or, more popularly, on training ML-based performance models~\cite{ernest, ottertune, cart, spark-trial, spark-tune-1, cdbtune, deeprm}. These models are trained either using a small-scale benchmark test bed, historical profiles, or from application performance under low workload. We argue that developing models that cater to changing workload or system environment is either impractical or potentially involves an expensive {\em online} learning cycle.

Black-box approaches are often employed to build an understanding of the interactions among configuration options on a newly seen workload. Many search-based approaches exist that use a combination of random sampling and local search~\cite{smarthill, mronline, elastisizer, bdaf, bestconfig}. However, such approaches are not suitable for our setup since there is a very high cost associated with running each experiment. Sequential Model-based Optimization (SMBO) approach~\cite{smac} helps speed up the exploration by using a surrogate model to fit existing observations and using it to recommend the next probe of configuration space. Bayesian optimization~\cite{bayesian-book} is a powerful state-of-the-art SMBO technique that has found applications in many system tuning problems~\cite{ituned, ottertune, bo4co, storage-tuning, cherrypick, arrow}. We adapt the Bayesian Optimization using Gaussian Process~\cite{gaussian} surrogate model for our problem setup. Alternate surrogate models such as Random Forest and Boosted Regression Trees have been shown to be better at modeling the non-linear interactions~\cite{arrow}. However, they lack theoretical guarantees on the confidence bounds that Gaussian Process offers. Also we did not find much qualitative difference among the models when evaluated in our setup and, therefore, do not include the results.

Guided Bayesian Optimization (\guided) we have developed is heavily motivated by Structured Bayesian Optimization (SBO)~\cite{boat} which lets the system developers add structure to the optimization by means of bespoke probabilistic models consisting of a non-parametric bayesian model and a set of evolving parametric models inferred from low-level performance metrics. In comparison, \guided\ simplifies the process with a white-box model that can be used from the beginning of the tuning process on any workload. Another recent work targeted at finding the best VM configurations~\cite{arrow} augments a bayesian optimizer with low-level performance metrics though without building any analytical models.

Reinforcement learning is a powerful AI technique which is being adapted by database researchers for traditional problems such as query optimization~\cite{neo} and database tuning~\cite{cdbtune, qtune}. While both CDBTune and QTune use \reinforce\ for database tuning, QTune adds a featurization step for SQL query workload to build models specific to the workload. We use \reinforce\ in a similar manner, though without using featurization, since our goal is to tune each application individually.

We have focussed at the memory management options in data analytics workloads. Most cloud-based deployments provide robust settings that are expected to generalize well across applications. As an example, Amazon's Elastic MapReduce (EMR) provides a default policy for resource allocation on Spark clusters, called {\em MaximizeResourceAllocation}~\cite{spark-emr}. We establish through a thorough empirical analysis that the framework defaults do not generalize well and leave a lot of scope for performance improvements, a fact also shown by others~\cite{ryza, thoth-action}. 
Like ours, there have been a few recent notable attempts at a systematic empirical analysis of data analytics systems. Charles Reiss~\cite{reiss} carried out an extensive evaluation of memory management in Spark and developed a tool to provision cluster memory to satisfy maximum memory requirements. Iorgulescu et.~al.~\cite{spilled} studied memory elasticity in Hadoop, Flink, Spark, and Tez frameworks and used it to improve cluster scheduling. Both papers analyze each memory pool individually unlike \relm\ which also considers the interactions amongst the pools at multiple levels. A direction we would like to work on in future is contributing to a self-driving cluster management platform. As an example, UnravelData is working on building a self-driving Spark cluster~\cite{unravel-cidr} by employing various AI techniques for tasks such as root cause analysis and systematic collection of monitoring data. Our automated memory tuners can contribute to such solutions working at industrial scale.


\vspace*{-2mm}
\section{Need for Database Performance Data Scientists}
\label{sec:conclusion}
\vspace*{-1mm}

In this paper, we studied the problem of autotuning the memory allocation for data analytics applications using a state-of-the-art, AI-driven, black-box approach and our new empirically-driven, white-box solution called \relm. We showed how \relm\ provides better quality results (in terms of the desired objectives of low wall-clock time and performance reliability) with minimal overheads. 
\relm's superior performance highlights that tuning algorithms developed by {\em Database Performance Data Scientists} who combine an understanding of the underlying database platform with the ability to develop data-driven algorithms must not be overlooked while building autonomous/self-driving data processing systems.

\eat{
\section{Conclusion}
\label{sec:conclusion}

We studied the problem of autotuning the memory allocation for data analytics applications using an AI-driven black-box algorithm and an empirically-driven white-box solution we developed. 
Our solution, \relm, provides better quality (in terms of the desirable objectives of low latency/ wall-clock duration and reliability of performance) of results with minimal overheads. \relm\ achieves these goals by developing a thorough understanding of interactions among parameters controlling the memory pools across multiple levels of memory management used in a data analytics application. 
We believe that our findings will help advance the research in self-driving data processing systems.
}


\bibliographystyle{ACM-Reference-Format}
\bibliography{refs}  


\end{document}